\documentclass{zyphra}

\usepackage{float}
\usepackage{pifont}
\usepackage{fontawesome5}
\usepackage{makecell}

\usepackage{acronym}
\usetikzlibrary{positioning,arrows.meta,fit,backgrounds,calc,shadows,shapes.geometric,decorations.pathreplacing,matrix,shapes.misc}

\definecolor{zorange}{HTML}{F28E2B}
\definecolor{znavy}{HTML}{1B3A5C}
\definecolor{zblue}{HTML}{4E79A7}
\definecolor{zgreen}{HTML}{8DA86B}
\definecolor{zyellow}{HTML}{E6B84F}
\definecolor{zred}{HTML}{C8705C}
\definecolor{zbrown}{HTML}{A98467}
\definecolor{zteal}{HTML}{59A5A8}
\definecolor{zpurple}{HTML}{8E7CC3}

\acrodef{TTS}{Text-to-Speech}
\acrodef{MoE}{Mixture of Experts}
\acrodef{RVQ}{Residual Vector Quantization}
\acrodef{DAC}{Descript Audio Codec}
\acrodef{ASR}{Automatic Speech Recognition}
\acrodef{VAD}{Voice Activity Detection}
\acrodef{WER}{Word Error Rate}
\acrodef{CER}{Character Error Rate}
\acrodef{MLP}{Multilayer Perceptron}
\acrodef{GQA}{Grouped Query Attention}
\acrodef{GDN}{Gated DeltaNet}
\acrodef{SWA}{Sliding Window Attention}
\acrodef{GLA}{Gated Linear Attention}
\acrodef{CFG}{Classifier-Free Guidance}
\acrodef{iSTFT}{inverse Short-Time Fourier Transform}
\acrodef{STFT}{Short-Time Fourier Transform}
\acrodef{XSA}{Exclusive Self-Attention}
\acrodef{CCA}{Compressed Convolutional Attention}
\acrodef{EDA}{Exponential Depth Averaging}
\acrodef{NLL}{Negative Log-Likelihood}
\acrodef{UTMOS}{UTokyo-SaruLab Mean Opinion Score}
\acrodef{FSDP}{Fully Sharded Data Parallel}
\acrodef{EP}{Expert Parallelism}
\acrodef{MFU}{Model FLOPs Utilization}
\acrodef{KV}{Key--Value}
\acrodef{LUFS}{Loudness Units relative to Full Scale}
\acrodef{RoPE}{Rotary Position Embedding}
\acrodef{SSL}{Self-Supervised Learning}
\acrodef{DS-WED}{Discretized Speech Weighted Edit Distance}
\acrodef{LDA}{Linear Discriminant Analysis}
\acrodef{QAT}{Quantization-Aware Training}
\acrodef{PTQ}{Post-Training Quantization}
\acrodef{LoRA}{Low-Rank Adaptation}
\acrodef{KLD}{Kullback--Leibler Divergence}
\acrodef{RTF}{Real-Time Factor}
\acrodef{CTC}{Connectionist Temporal Classification}
\acrodef{FSQ}{Finite Scalar Quantization}
\acrodef{LID}{Language Identification}
\acrodef{MPM}{Masked Prosody Model}
\acrodef{NoPE}{No Positional Encoding}
\acrodef{PESQ}{Perceptual Evaluation of Speech Quality}
\acrodef{STOI}{Short-Time Objective Intelligibility}
\acrodef{LSD}{Log-Spectral Distance}
\acrodef{SI-SDR}{Scale-Invariant Signal-to-Distortion Ratio}
\acrodef{ZAC-1}{Zyphra Audio Codec 1}
\acrodef{QB}{Quantile Balancing}

\title{ZONOS3-Mini Technical Report}
\shorttitle{ZONOS3-Mini Technical Report}
\reporttype{Technical report}
\author{Gabriel Clark\corr\qquad Sofian Mejjoute\qquad George Close\qquad Beren Millidge\corr}
\affiliation{Zyphra\enspace\textperiodcentered\enspace San Francisco, CA}
\authornote{\corr Corresponding authors}
\pdfauthors{Gabriel Clark, Sofian Mejjoute, George Close, Beren Millidge}
\keywords{text-to-speech, voice cloning, audio codec, mixture of experts, on-device inference}

\metadata{Model}{\zyhf{Zyphra/ZONOS3-Mini}}
\metadata{Code}{\zygh{Zyphra/ZONOS3-Mini}}
\metadata{Evaluation}{\zygh{Zyphra/ZTTS2-Eval}}
\metadata{Contact}{\href{mailto:gabriel@zyphra.com}{gabriel@zyphra.com}\quad
  \href{mailto:beren@zyphra.com}{beren@zyphra.com}}

\begin{document}

\maketitle

\begin{abstract}
  ZONOS3-Mini is an ultra-compact, ultra-fast, zero-shot voice-cloning TTS model. The model can run in real time on a laptop or smartphone. For its latency and size, it achieves state-of-the-art accuracy and cloning fidelity. To achieve this we propose a latent MoE backbone with only 886M parameters (100M active per token) and a new audio codec, \ac{ZAC-1}, which represents 44.1\,kHz audio at a 15\,Hz frame rate. On the new ZTTS2-Eval benchmark, ZONOS3-Mini reaches a 6.71\,\% macro-average WER over 55 language sets (35 clean and 20 in the wild), which is competitive with larger and slower state-of-the-art models. We release the model weights, inference, and training code.
\end{abstract}

\zycontents
\clearpage

\begin{figure}[H]
  \centering
  \includegraphics[width=\linewidth]{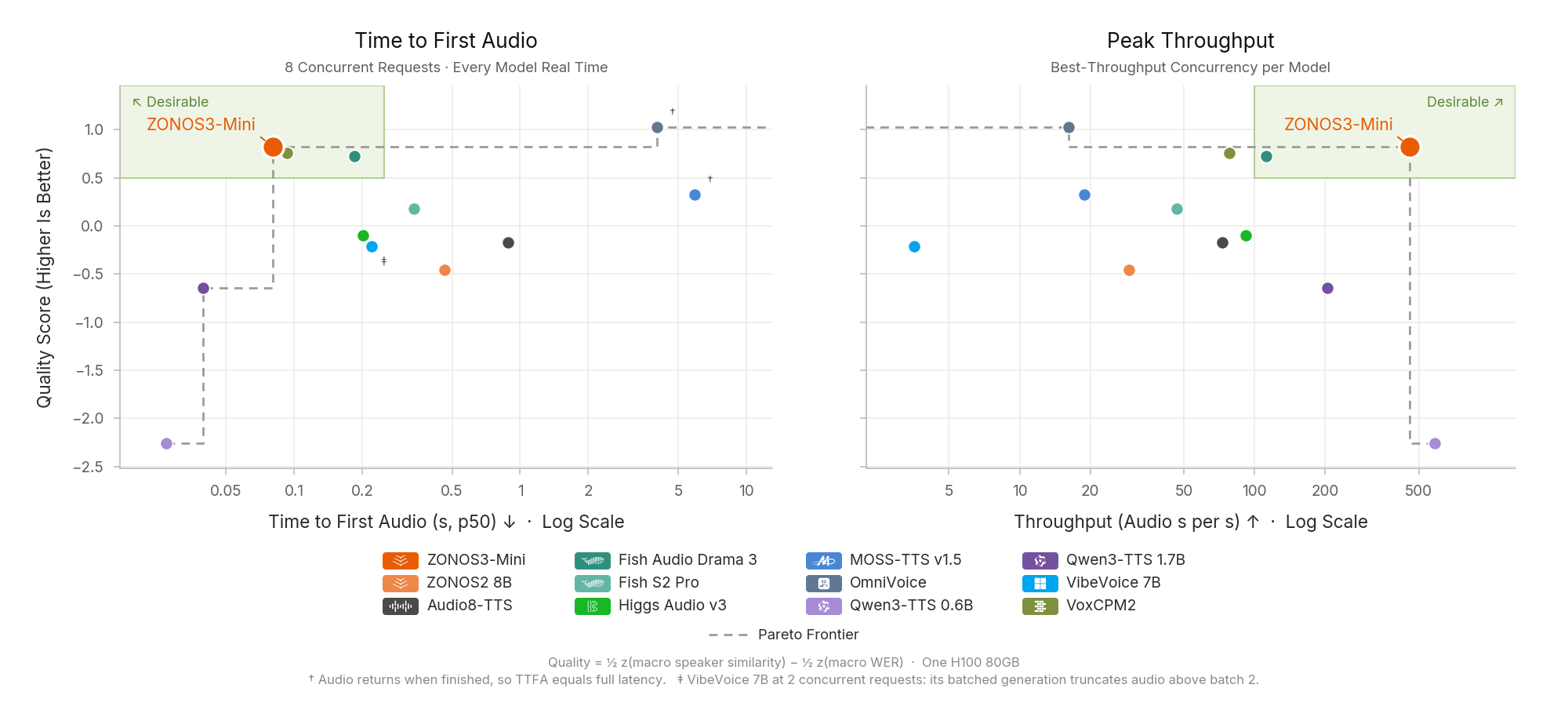}
  \caption{Quality versus speed on one H100 80GB. Left: time to first audio (p50) at 8 concurrent
    requests; right: peak throughput at each model's best-throughput concurrency. Quality is
    $\tfrac{1}{2}$\,z(macro speaker similarity) $-$ $\tfrac{1}{2}$\,z(macro WER) over the 55 ZTTS2-Eval
    language sets; the dashed line is the Pareto frontier.}
  \label{fig:headline}
\end{figure}

\section{Introduction}

Current state-of-the-art on-device TTS models sacrifice cloning quality, controllability, and multilingual performance for the ability to operate in real time on weaker hardware. We introduce ZONOS3-Mini as a new point on the Pareto frontier of quality and inference performance for on-device TTS models. In this report we detail the architecture and training framework that allow ZONOS3-Mini to operate everywhere from smartphones to laptops, without compromising on quality. The main areas of advancement over ZONOS2 are in sequence compression, text and audio representations, model architecture, voice cloning conditioning, controllability, and evaluation methodology. Across many of these areas of improvement there were two main drivers. The first and most effective change was switching to our new audio tokenizer \ac{ZAC-1} and the second was validating all model changes with our new evaluation, ZTTS2-Eval.

ZAC-1 operates at 15\,Hz, which shortens audio sequences by $5.7\times$ relative to ZONOS2~\citep{clark2026zonos2}, and it delivers high-quality reconstructions at only 1.80\,kbit/s. It adopts a dual-encoding~\citep{li2025dualcodec} quantizer design with semantic distillation from frozen self-supervised speech representations. This enables more efficient training and a backbone with roughly a ninth of the parameters of ZONOS2, with higher speaker similarity and a macro \ac{WER} within 0.5 points of ZONOS2 (\Cref{tab:external}). The use of causal \ac{iSTFT} heads and a novel transformer autoencoder design enables low-latency streaming and efficient inference.

We also report results on and release ZTTS2-Eval, a new \ac{TTS} evaluation framework covering 35 languages in the clean subset and 20 in the in-the-wild set, totaling 74,041 utterances. This is significantly wider in language coverage than existing \ac{TTS} evaluation sets.

\section{Backbone Overview}\label{sect:overview}

\subsection{Text Tokenization}
In a departure from ZONOS2, which used raw byte text representations, ZONOS3-Mini uses an 8,192-entry SuperBPE  tokenizer~\citep{liu2025superbpe} trained on a spoken-language-focused data mix. The text span is bracketed by a beginning-of-text token (BOS) and an end-of-text token (EOS); the audio stream ends with an end-of-audio token (EOA), and a pad row fills unused positions. The vocabulary size follows the tokenizer sweep of \Cref{sec:tok_ablation}.

\subsection{Backbone Architecture}
The backbone combines recurrent \acf{GDN} layers with two global attention layers for efficient long-context inference. It has 11 layers in total. Layers 4 and 8 use global causal attention with \acf{GQA}, QK-normalization~\citep{henry-etal-2020-query}, \ac{RoPE}~\citep{SU2024127063} and a learned output gate per head~\citep[\zypin{\S2.2}{2505.06708v1}{3}]{qwen-gating}. These layers have 16 query heads, four key-value heads and head dimension 32. The other nine layers use \ac{GDN} mixers~\citep{yang2024gateddeltanet} with four heads and key/value dimension 128. The model uses a latent \ac{MoE}, reducing the 1,024-dimensional model width to 256 inside the experts. We use full attention residuals~\citep[\zypin{\S3.1}{2603.15031v1}{4}]{kimi2026attnres} as their memory I/O remains small in our 11-layer model.

\subsection{Speaker Conditioning}\label{sec:speaker}
ZONOS3-Mini supports cloning from both `raw-audio' codec references and from speaker embeddings. Raw references carry prosody and recording conditions as well as speaker identity. Embeddings occupy a single sequence position and score higher on speaker similarity in our evaluation (\Cref{sect:results}).

We train our own speaker embedding encoder with the MECT-B2 architecture~\citep{antspeaker2026mect}: a convolutional encoder followed by a \ac{MoE} transformer with an attentive statistics pooling stage. It is trained independently of the backbone with the SphereFace2 loss~\citep{wen2022sphereface2}. It produces a 192-dimensional vector, which a projection maps to the model's hidden dimension.

Pre-training uses the codec codes of a 1--60\,s reference clip as a loss-masked prefix before the text. The reference comes from another clip of the same speaker in the same recording selected through diarization consensus (\Cref{sect:data}). We omit it in 10\,\% of training examples to enable unconditioned generation and speaker \ac{CFG}.
Fine-tuning introduces the embedding path together with quality and speaking-rate controls, while retaining examples with raw references. 

\section{Audio Codec: ZAC-1}\label{sect:codec}

\begin{figure}[t]
  \centering
  \resizebox{\linewidth}{!}{
\definecolor{zcWire}{HTML}{3A2F26}
\definecolor{zcInk}{HTML}{2A211A}
\definecolor{zcGray}{HTML}{555555}
\definecolor{zcBlockF}{HTML}{F6DCC0}\definecolor{zcBlockS}{HTML}{DB9A59}
\definecolor{zcPinkF}{HTML}{F1D1C6}\definecolor{zcPinkS}{HTML}{CB7860}
\definecolor{zcYelF}{HTML}{F4E8BD}\definecolor{zcYelS}{HTML}{C6A449}
\definecolor{zcLavF}{HTML}{E6D8C0}\definecolor{zcLavS}{HTML}{A78960}
\definecolor{zcCrmF}{HTML}{FCF6E8}\definecolor{zcCrmS}{HTML}{D7C096}
\definecolor{zcBeigeF}{HTML}{FCF3E7}\definecolor{zcBeigeS}{HTML}{E9D3BA}
\definecolor{zcIoF}{HTML}{E9F1F8}\definecolor{zcIoS}{HTML}{9FB8CF}
\definecolor{zcIce}{HTML}{5F8BB5}\definecolor{zcIceF}{HTML}{CFE0EF}
\definecolor{zcTok}{HTML}{FCB702}
\begin{tikzpicture}[x=1mm, y=1mm, font=\scriptsize, text=zcInk, line cap=round, line join=round,
    blk/.style={draw=zcBlockS, fill=zcBlockF, line width=0.5pt, rounded corners=3pt,
                minimum height=7.5mm, minimum width=16mm, inner sep=1.5pt, align=center,
                drop shadow={shadow xshift=0pt, shadow yshift=-0.45pt, opacity=0.12, fill=black}},
    stg/.style={blk, minimum width=5.2mm, inner sep=0pt},
    vblk/.style={blk, minimum width=7mm, inner sep=0pt},
    sem/.style={blk, draw=zcYelS, fill=zcYelF},
    frz/.style={blk, draw=zcCrmS, fill=zcCrmF},
    qnt/.style={blk, draw=zcLavS, fill=zcLavF},
    head/.style={blk, draw=zcPinkS, fill=zcPinkF, minimum width=7mm, minimum height=28mm, inner sep=0pt},
    cnxv/.style={sem, minimum width=7mm, minimum height=28mm, inner sep=0pt},
    pill/.style={blk, draw=zcIoS, fill=zcIoF, rounded corners=3.5mm, minimum width=7mm, minimum height=28mm, inner sep=0pt},
    tok/.style={draw=zcWire, line width=0.4pt, rounded corners=0.7pt, minimum size=3mm, inner sep=0pt, font=\tiny},
    vt/.style={rotate=90, align=center, inner sep=0pt},
    op/.style={circle, draw=zcWire, line width=0.55pt, fill=white, inner sep=0pt, minimum size=3.8mm},
    wire/.style={draw=zcWire, line width=0.55pt},
    arr/.style={wire, -{Triangle[length=1.4mm, width=1.2mm]}},
    cont/.style={draw=zcBeigeS, fill=zcBeigeF, line width=0.5pt, rounded corners=2.5pt},
    grp/.style={draw=zcGray, line width=0.5pt, dash pattern=on 3pt off 2pt, rounded corners=6pt, inner sep=2mm},
    glbl/.style={font=\scriptsize, text=zcGray, inner sep=1pt, align=center},
    note/.style={font=\fontsize{6}{7}\selectfont, text=zcGray, inner sep=0.8pt, align=center},
    wavev/.pic={\foreach \h [count=\i] in {0.4,0.9,0.6,1.3,0.8,1.6,1.1,0.7,1.5,1.0,1.8,1.2,0.6,1.4,0.9,1.7,1.0,0.5,1.3,0.8,1.1,0.6,0.9,0.4}{
        \draw[zcInk, line width=0.45mm, line cap=round] (-\h,{0.72*(\i-12.5)}) -- (\h,{0.72*(\i-12.5)});}},
    plusg/.pic={\draw[zcWire, line width=0.55pt, line cap=butt] (-0.9,0) -- (0.9,0) (0,-0.9) -- (0,0.9);},
    minusg/.pic={\draw[zcWire, line width=0.55pt, line cap=butt] (-0.9,0) -- (0.9,0);}
]
    \node[pill] (pin) at (4.5,0) {};   \pic at (pin.center) {wavev};
    \node[glbl, text=zcInk, below=1mm of pin.south] {Audio\\[-1pt]44.1\,kHz};
    \node[vblk, minimum height=20mm] (stem)  at (15,0) {};
    \node[vt] at (stem.center) {\textbf{Conv stem}\\[-1pt]256 ch, $\div$8};
    \node[vblk, minimum height=15mm] (patch) at (25.5,0) {};
    \node[vt] at (patch.center) {\textbf{Patch}\\[-1pt]$\times$23};
    \begin{scope}[on background layer]
        \draw[cont] (32.5,-13) -- (68.5,-8) -- (68.5,8) -- (32.5,13) -- cycle;
        \draw[cont] (115.1,-8) -- (151.1,-13) -- (151.1,13) -- (115.1,8) -- cycle;
    \end{scope}
    \foreach \x/\hh/\lab/\nm in {37.1/22/240/e240, 43.8/19.5/120/e120, 50.5/17/60/e60, 57.2/14.5/30/e30, 63.9/12/15/e15}{
        \node[stg, minimum height=\hh mm] (\nm) at (\x,0) {};
        \node[vt] at (\nm.center) {\textbf{\lab\,Hz}};}
    \node[op] (minus) at (79.4,0) {};  \pic at (minus.center) {minusg};
    \node[qnt, minimum width=13mm] (rvq) at (91.3,0) {\textbf{RVQ}\\[-1pt]9 codebooks};
    \node[tok, draw=zcIce, fill=zcIceF] (q1) at (102.8,6.5) {$q_1$};
    \node[tok, fill=zcTok] (q2)  at (102.8,3.25) {$q_2$};
    \node[tok, fill=zcTok] (q3)  at (102.8,0) {$q_3$};
    \foreach \dy in {-0.85,0,0.85}{\fill[zcWire] (102.8,-3.25+\dy) circle (0.22);}
    \node[tok, fill=zcTok] (q10) at (102.8,-6.5) {$q_{10}$};
    \node[note, below=0.8mm of q10.south] {1 semantic + 9 residual codes\\[-1pt]per 15\,Hz frame};
    \node[op] (plus) at (109.7,0) {};  \pic at (plus.center) {plusg};
    \foreach \x/\hh/\lab/\nm in {119.7/12/15/d15, 126.4/14.5/30/d30, 133.1/17/60/d60, 139.8/19.5/120/d120, 146.5/22/240/d240}{
        \node[stg, minimum height=\hh mm] (\nm) at (\x,0) {};
        \node[vt] at (\nm.center) {\textbf{\lab\,Hz}};}
    \node[cnxv] (cnx) at (158.1,0) {};
    \node[vt] at (cnx.center) {\textbf{8 ConvNeXt blocks}};
    \node[head] (istft) at (168.6,0) {};
    \node[vt] at (istft.center) {\textbf{iSTFT head}};
    \node[pill] (pout) at (179.1,0) {};   \pic at (pout.center) {wavev};
    \node[glbl, text=zcInk, below=1mm of pout.south] {Audio\\[-1pt]44.1\,kHz};
    \draw[arr] (pin.east) -- (stem.west);
    \draw[arr] (stem.east) -- (patch.west);
    \draw[arr] (patch.east) -- (32.5,0);
    \draw[arr] (68.5,0) -- node[note, above, pos=0.5] {1,024-d} (minus.west);
    \draw[arr] (minus.east) -- (rvq.west);
    \draw[arr] (rvq.east) -- (q3.west);
    \draw[arr] (q3.east) -- (plus.west);
    \draw[arr] (plus.east) -- (115.1,0);
    \draw[arr] (151.1,0) -- (cnx.west);
    \draw[arr] (cnx.east) -- (istft.west);
    \draw[arr] (istft.east) -- (pout.west);
    \node[glbl] at (50.5,-15.5) {Encoder};
    \node[glbl] at (133.1,-15.5) {Decoder};

    \node[frz, minimum width=23mm] (w2v)   at (30,27)    {\textbf{w2v-BERT 2.0} \textcolor{zcIce}{\faSnowflake}\\[-1pt]layer 16};
    \node[sem, minimum width=16mm] (cenc)  at (53,27)    {\textbf{ConvNeXt}\\[-1pt]4 blocks};
    \node[qnt, minimum width=19mm] (vq)    at (74,27)    {\textbf{Semantic VQ}};
    \node[tok, draw=zcIce, fill=zcIceF] (sq1) at (88.5,27) {$q_1$};
    \node[sem, minimum width=22mm] (chead) at (104.5,27) {\textbf{Semantic head}\\[-1pt]4 ConvNeXt};
    \draw[arr] (pin.north) |- (w2v.west);
    \draw[arr] (w2v) -- (cenc);
    \draw[arr] (cenc) -- (vq);
    \draw[arr] (vq) -- (sq1);
    \draw[arr] (sq1) -- (chead);
    \draw[wire] (chead.south) -- (chead.south |- 0,15.5);
    \draw[arr] (chead.south |- 0,15.5) -| (minus.north);
    \draw[arr] (chead.south |- 0,15.5) -| (plus.north);
    \begin{scope}[on background layer]
        \node[grp, fit=(w2v)(chead)] (semgrp) {};
    \end{scope}
    \node[glbl, below=0.5mm of semgrp.south] {Semantic branch};
\end{tikzpicture}}
  \caption{ZAC-1. One semantic and nine residual codes per 15\,Hz frame are passed to the
    backbone.}
  \label{fig:zac}
\end{figure}
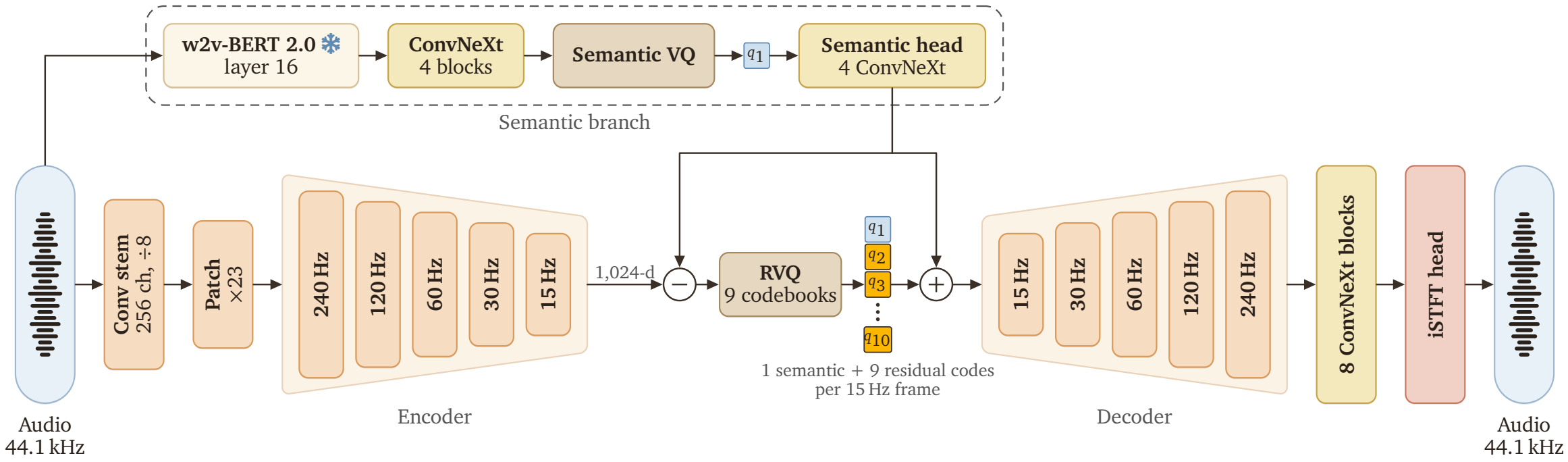

\begin{table}[t]
  \centering
  \small
  \setlength{\tabcolsep}{6pt}
  \begin{tabular}{lrrrrr}
    \toprule
    & & \multicolumn{2}{c}{\textbf{WER $\downarrow$}} & \multicolumn{2}{c}{\textbf{Sim $\uparrow$}} \\
    \cmidrule(lr){3-4}\cmidrule(lr){5-6}
    \textbf{Codec} & \textbf{Frames/s} & clean & wild & clean & wild \\
    \midrule
    Fish S1--DAC & 21.5 & 4.30 & \textbf{5.28} & \textbf{84.8} & 79.0 \\
    ZAC-1 & 15.0 & \textbf{4.06} & 5.54 & 83.8 & 76.1 \\
    MOSS, 14 \acs{RVQ} & 12.5 & 5.06 & 5.67 & 84.6 & \textbf{80.3} \\
    \acs{DAC} & 86.1 & 42.2 & 51.7 & 78.8 & 63.7 \\
    \bottomrule
  \end{tabular}
  \caption{Downstream codec comparison:  50k-step \ac{TTS} models on the four-language
ZTTS2-Eval sets. Best in bold.}
  \label{tab:codec_cmp}
\end{table}

Our goal in designing ZAC-1 was to produce a codec that is fast to decode and easy for the \ac{TTS} backbone to model in a delay pattern, without compromising on acoustic quality. ZAC-1 has a 1.99B-parameter encoder-decoder structure and operates at 15\,Hz and 1.80\,kbit/s, as shown in \Cref{fig:zac}. It uses DualCodec-style dual encoding~\citep[\zypin{\S3}{2505.13000v2}{2}]{li2025dualcodec} with frozen layer-16 w2v-BERT 2.0 features~\citep{chung2021w2vbert,seamless2023m4t} as the semantic representation to improve multilingual modelability in the downstream \ac{TTS} backbone. Each frame carries ten codes from 4,096-entry codebooks, one semantic code and nine \ac{RVQ} residual codes, a configuration that balances speaker similarity and \ac{WER} in delay-pattern \ac{TTS} modeling. The encoder and decoder comprise a symmetric cascade of causal local-attention transformer stages at 15, 30, 60, 120 and 240\,Hz. The encoder converts audio into patches with DAC-like residual units, and the decoder uses eight ConvNeXt blocks and two \ac{iSTFT} heads to generate 44.1\,kHz audio; both are designed to avoid running attention at the 44.1\,kHz sample rate. ZAC-1 was trained for 600k steps with 10\,s crops and mel, adversarial, feature-matching and quantizer losses.

To compare codecs, we encoded a 2\,\% stratified subset of the pre-training corpus (\Cref{sect:data}) and trained a copy of our \ac{TTS} model for 50k steps on this data. \Cref{tab:codec_cmp} compares the final checkpoint of ZAC-1 with the public Fish S1--DAC~\citep{liao2026fishaudios2technical}, MOSS audio tokenizer v2~\citep{openmoss2026mossaudiotokenizer}, and \ac{DAC} checkpoints. We find that ZAC-1 provides competitive performance at a desirable compression rate. 

The decoder ends in a Vocos-style \ac{iSTFT} head~\citep[\zypin{\S3.2}{2306.00814v3}{4}]{siuzdak2023vocos} for inference performance. Originally we tried a symmetrical design for the encoder and decoder, which achieved superior reconstruction metrics; however, during inference at large batch sizes, moving waveform-rate activations through memory was prohibitively expensive. Because batch-size-1 local inference and large-batch cloud inference have different optima, two decoders are released. The full decoder removes the large-batch memory bottleneck of the last layer by replacing it with ConvNeXt blocks and an \ac{iSTFT} head, and the pruned 144M-parameter keep-2 decoder avoids the batch-size-1 FLOP bottleneck of on-device inference by removing the transformer blocks that hold the most FLOPs. In both decoders, eight causal ConvNeXt blocks at the 240\,Hz stage produce log-magnitude and phase spectra for two \ac{iSTFT} heads at different resolutions. In decode-only throughput on one H100 the full decoder runs at 1,180$\times$ real time at batch 256 while the keep-2 decoder runs at 3,950$\times$, compared with 989$\times$ for the MOSS audio tokenizer~\citep{openmoss2026mossaudiotokenizer} and 474$\times$ for S1--\ac{DAC}.

\begin{table}[t]
  \centering
  \footnotesize
  \setlength{\tabcolsep}{4pt}
  \begin{tabular}{lrrrrrrrrr}
    \toprule
    \textbf{Codec} & \textbf{Frames/s} & \textbf{kbit/s} & \acs{PESQ} $\uparrow$ & \acs{STOI} $\uparrow$ & Mel $\downarrow$ & \acs{LSD} $\downarrow$ & \acs{SI-SDR} $\uparrow$ & Sim $\uparrow$ & UTMOS $\uparrow$ \\
    \midrule
    \multicolumn{10}{l}{\emph{ZTTS2-Eval ground truth, 1,600 clips, 16\,kHz (original audio UTMOS 2.80)}} \\
    ZAC-1, full decoder & 15 & 1.80 & 2.88 & 0.928 & 0.179 & 0.955 & 7.83 & 88.7 & 2.63 \\
    ZAC-1, keep-2 decoder & 15 & 1.80 & 2.51 & 0.911 & 0.202 & 1.005 & 5.72 & 84.9 & 2.52 \\
    Fish S1--\acs{DAC} & 21.5 & 2.20 & 3.10 & 0.941 & 0.149 & \underline{0.900} & 7.25 & 90.7 & 2.66 \\
    \acs{DAC} & 86.1 & 7.75 & \textbf{3.89} & \textbf{0.964} & \textbf{0.132} & 0.926 & \textbf{12.94} & \underline{93.8} & \textbf{2.71} \\
    MOSS tokenizer, 14 \acs{RVQ} & 12.5 & 1.75 & 3.13 & 0.940 & 0.152 & 0.911 & 8.81 & 92.3 & \underline{2.70} \\
    MOSS tokenizer, 32 \acs{RVQ} & 12.5 & 4.00 & \underline{3.55} & \underline{0.962} & \underline{0.133} & \textbf{0.887} & \underline{11.22} & \textbf{95.5} & 2.68 \\
    \midrule
    \multicolumn{10}{l}{\emph{DAPS clean, 200 utterances, 44.1\,kHz  (original audio UTMOS 3.69)}} \\
    ZAC-1, full decoder & 15 & 1.80 & -- & -- & 0.209 & 1.212 & 8.56 & 84.0 & \underline{3.70} \\
    ZAC-1, keep-2 decoder & 15 & 1.80 & -- & -- & 0.227 & 1.229 & 6.55 & 78.2 & 3.61 \\
    Fish S1--\acs{DAC} & 21.5 & 2.20 & -- & -- & \underline{0.145} & \textbf{0.822} & 8.62 & 88.6 & \textbf{3.72} \\
    \acs{DAC} & 86.1 & 7.75 & -- & -- & \textbf{0.137} & \underline{0.832} & \textbf{12.71} & \underline{92.4} & 3.68 \\
    MOSS tokenizer, 14 \acs{RVQ} & 12.5 & 1.75 & -- & -- & 0.331 & 2.684 & 9.52 & 89.9 & 3.67 \\
    MOSS tokenizer, 32 \acs{RVQ} & 12.5 & 4.00 & -- & -- & 0.315 & 2.674 & \underline{11.78} & \textbf{94.2} & 3.66 \\
    \bottomrule
  \end{tabular}
  \caption{Codec reconstruction quality on ZTTS2-Eval clips and DAPS clean. Mel is the log-mel L1 distance. Best in bold, second best underlined.}
  \label{tab:codec_recon}
\end{table}

For completeness, we report reconstruction quality in \Cref{tab:codec_recon} against public Fish
S1--DAC~\citep{liao2026fishaudios2technical}, \ac{DAC}~\citep{kumar2023highfidelityaudiocompressionimproved}, and the MOSS audio tokenizer~\citep{openmoss2026mossaudiotokenizer}. The audio used is a subset of ZTTS2-Eval and the clean partition of DAPS~\citep{mysore2015daps}, neither of which is included in our training corpus. Reconstructions are scored with \ac{PESQ}~\citep{rix2001pesq}, \ac{STOI}~\citep{taal2011stoi}, log-mel distance, \ac{LSD}~\citep{gray1976lsd}, \ac{SI-SDR}~\citep{leroux2019sisdr}, ReDimNet speaker similarity~\citep{Yakovlev_2024} and MSR-UTMOS~\citep{nishikawa2025multisamplingfrequencynaturalnessmosprediction}. The MOSS tokenizer has a variable bit rate and is shown in two modes here. The keep-2 decoder trades 0.4 \ac{PESQ} and 4 similarity points against the full decoder.

\section{Data Pipeline}\label{sect:data}
Our pre-training corpus consists of 6.1M hours of naturalistic speech audio sourced from a mixture of public and private corpora. We reuse all of the labels and data from the ZONOS2 data pipeline. To obtain utterance pairs of a speaker from different sections of a recording, we diarize each recording with both HDBSCAN~\citep{campello2013hdbscan} and spectral clustering over speaker-embedding windows and only use intervals where both methods agree upon the same speaker.

\section{Training}\label{sect:training}

\subsection{Pre-training}
\begin{figure}[t]
  \centering
  \includegraphics[width=0.95\linewidth]{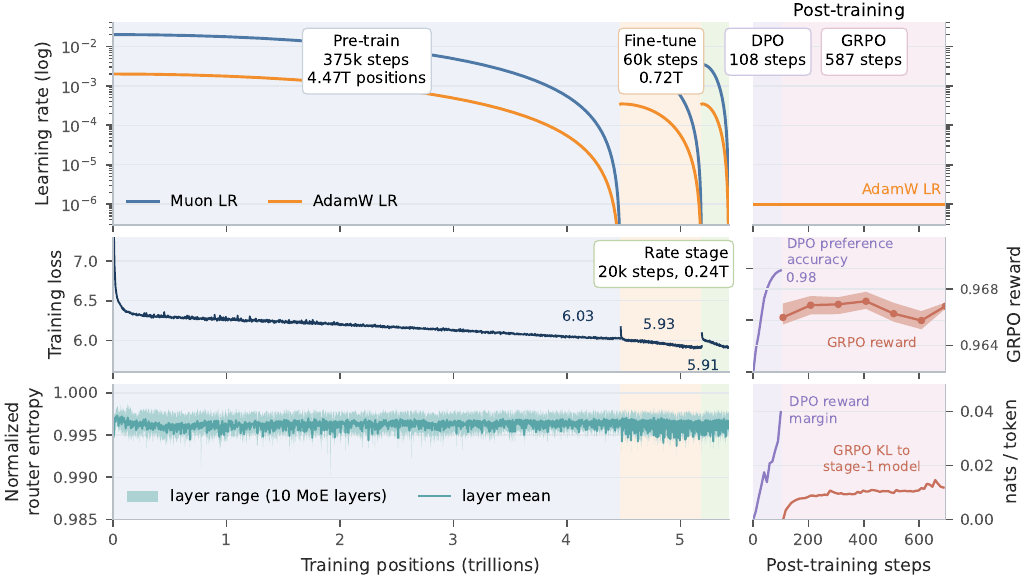}
  \caption{Learning rate, training loss and normalized router entropy over pre-training, fine-tuning, the speaking-rate stage and
post-training (DPO and GRPO). The fine-tune loss starts higher due to the addition of speaker-projection and
    control-row tables and a change to the data mix. Router entropy is
  the mean and range over the ten \ac{MoE} layers.}
  \label{fig:loss}
\end{figure}

ZONOS3-Mini was pre-trained for 375k steps on H100 GPUs for a total of 4.5T positions or ten epochs over the full corpus. We trained using the hyper-parameters in \Cref{tab:hparams}. Training sample lengths were up to 600\,s and the audio reference clips for voice cloning were up to 60\,s. We used the Muon optimizer~\citep{jordan6muon,moonshot-muon} for 2-D weights and AdamW for the rest. After adopting \ac{QB}, router entropy stayed within 5\,\% of uniform in every \ac{MoE} layer for all of training (\Cref{fig:loss}).

\subsection{Fine-tuning}\label{sec:ft_series}
During fine-tuning we keep cloning from codec-encoded reference clips while additionally learning to clone from a MECT-B2-based speaker embedding. We fine-tune the 375k checkpoint for 60k steps with fresh optimizer states and the schedule in \Cref{tab:hparams}. On 45\,\% of training samples we provide the embedding to the model, on 45\,\% we use codec encoded reference clips, and on the remainder we provide no voice cloning reference. A further filtered subset of the data mix was used for this stage by increasing transcript agreement to 3 sources and lowering the CER-agreement threshold to 2\,\%. After initial evaluation we found that ten speaking-rate buckets were too coarse, so we performed a second fine-tuning stage of 20k steps to refine the speaking-rate control further. In this stage we split each of the ten speaking-rate buckets into three finer-grained buckets and add them as tokens which start initialized from their parents' representations. This stage restarts the optimizer states and the schedule but keeps the previous stage's data mix.

\subsection{Post-training}\label{sec:dpo}
We then perform two stages of post-training. Both stages use 5,089 training utterances in English, Chinese, Spanish and Japanese. In both stages, each generated candidate is scored with the same \ac{ASR} and speaker-similarity models as our evaluation (\Cref{sect:eval}): Qwen3-ASR against the prompt text and ReDimNet against the prompt audio.

Our first post-training stage targets early stopping and runaway generation failures with direct preference optimization
(DPO)~\citep[\zypin{Eq.~7}{2305.18290v3}{4}]{rafailov2024direct}. Eleven candidates per
prompt are generated, eight at speaker \ac{CFG} 1 and three at 1.5. A candidate is considered a failure when
it contains no speech, reaches the default inference token cap, is significantly longer or shorter than the source audio clip, or has a high \ac{WER}. Each failure forms a
pair with the most similar complete candidate of the same prompt. The loss is the DPO objective on the mean log-probability of the audio tokens
with $\beta = 0.1$, plus 0.05 times the negative log-likelihood of the chosen
sequence. Training uses AdamW at a learning rate of $10^{-6}$ with 8 pairs per step for 4 epochs.

The second post-training stage applies group relative policy optimization
(GRPO)~\citep[\zypin{\S4.1}{2402.03300v3}{11}]{shao2024deepseekmath}, following the recipe of
Audio8-TTS~\citep[\zypin{\S4.3}{2610.14776v1}{5}]{audio8tts2026} to directly optimize speaker similarity (SIM) and \ac{WER}. The first-stage model generates eight candidates per
prompt at \ac{CFG} 1. The reward of a candidate is
\begin{equation}
0.9\max(0,\,1-\frac{\mathrm{WER}}{2}) + 0.1\,\frac{\mathrm{SIM}}{100}.
\end{equation}
The advantage is the reward standardized within the prompt's group, minus $0.5$ times a duration cost. The cost is
\begin{equation}
    \max(0,\,\log(\frac{r}{1.35}))
\end{equation} standardized within the group,
where $r$ is the duration ratio of the generation to the original recording. Groups whose reward
standard deviation is below $0.005$ are dropped, which leaves 2,348 groups. The loss is
the clipped token-level surrogate with $\epsilon = 0.2$ plus a \ac{KLD} penalty of
weight 0.04 to the first-stage model, with log-ratios clamped to $\pm 5$. Training
uses AdamW at a learning rate of $10^{-6}$ with 4 groups per step and frozen expert
routers, for one epoch.
\section{Evaluation}\label{sect:eval}
ZTTS2-Eval succeeds the ZTTS1-Eval benchmark introduced with ZONOS2~\citep{clark2026zonos2}. It includes over 74,000 utterances across a clean set and an in-the-wild set. The clean set uses FLEURS-R read speech~\citep{ma2024fleursrrestoredmultilingualspeech} and covers 35 languages. The in-the-wild set uses a VoxBlink2 subsample~\citep{lin2024voxblink2100kspeakerrecognition} and covers 20 languages. Prompt and target come from different audio clips of the same speaker, such that similarity is measured across recordings.

We score generations with Qwen3-ASR~\citep{Qwen3-ASR}, except for English, which uses Whisper-large-v3. Speaker similarity is measured with ReDimNet~\citep{Yakovlev_2024} and quality with MSR-UTMOS~\citep{nishikawa2025multisamplingfrequencynaturalnessmosprediction}. We additionally evaluate rhythm and pitch distribution matching, together with a modified \ac{MPM}~\citep{wallbridge25_interspeech} embedding distance that uses a scrambled-input floor. We report \ac{WER} for most languages, with \ac{CER} for Chinese and Japanese. Per-utterance error rates are capped at 100\,\% before averaging. Similarity is the ReDimNet cosine similarity on a 0 to 100 scale.
The results use all 55 parts, 35 clean and 20 in the wild. ``Macro'' is the unweighted mean of the parts.

\section{Results}\label{sect:results}
\Cref{tab:results_main} shows the results of the fine-tuned model and of the post-trained
model on both prompt types.
\begin{table}[t]
  \centering
  \footnotesize
  \setlength{\tabcolsep}{6pt}
  \begin{tabular}{llrrrrrr}
    \toprule
    & & \multicolumn{2}{c}{\makecell{\textbf{Clean}\\(35 languages)}} & \multicolumn{2}{c}{\makecell{\textbf{In the wild}\\(20 languages)}} & \multicolumn{2}{c}{\textbf{Macro}} \\
    \cmidrule(lr){3-4}\cmidrule(lr){5-6}\cmidrule(lr){7-8}
    \textbf{Model} & \textbf{Conditioning} & WER $\downarrow$ & Sim $\uparrow$ & WER $\downarrow$ & Sim $\uparrow$ & WER $\downarrow$ & Sim $\uparrow$ \\
    \midrule
    ZONOS3-Mini & embedding & 8.28 & \underline{83.1} & 7.52 & \underline{74.0} & 8.00 & \underline{79.8} \\
    ZONOS3-Mini & raw audio & 7.95 & 77.7 & 12.57 & 55.1 & 9.63 & 69.5 \\
    \midrule
    ZONOS3-Mini + PT & embedding & \underline{7.79} & \textbf{83.4} & \textbf{4.81} & \textbf{77.3} & \underline{6.71} & \textbf{81.2} \\
    ZONOS3-Mini + PT & raw audio & \textbf{7.39} & 76.5 & \underline{5.48} & 58.1 & \textbf{6.70} & 69.8 \\
    \bottomrule
  \end{tabular}
  \caption{ZTTS2-Eval results at \ac{CFG} 1 over all 55 parts. Best in bold, second best underlined.}
  \label{tab:results_main}
\end{table}
The fine-tuned model scores 79.8 similarity and 8.00 \ac{WER} with embeddings,
compared with 69.5 and 9.63 with raw references. On the clean sets the two prompt
types have similar \ac{WER}. The gap comes from the in-the-wild sets, where the
raw-audio path has a \ac{WER} of 12.57 against 7.52 for embeddings. The embedding
path has higher speaker similarity, in particular in the wild (74.0 against 55.1).
Post-training lowers the macro \ac{WER} of the embedding path from 8.00 to 6.71 and
raises similarity from 79.8 to 81.2. The \ac{WER} reduction comes from the in-the-wild
sets, where it falls from 7.52 to 4.81 and similarity rises from 74.0 to 77.3. On
the clean sets \ac{WER} falls from 8.28 to 7.79. The raw-audio path improves from 9.63
to 6.70 \ac{WER} with similarity unchanged (69.5 to 69.8). Its in-the-wild \ac{WER}
falls from 12.57 to 5.48.

\begin{table}[t]
  \centering
  \footnotesize
  \setlength{\tabcolsep}{4pt}
  \begin{tabular}{llrrrrrr}
    \toprule
    & \textbf{Params} & \multicolumn{2}{c}{\textbf{Clean}} & \multicolumn{2}{c}{\textbf{In the wild}} & \multicolumn{2}{c}{\textbf{Macro}} \\
    \cmidrule(lr){3-4}\cmidrule(lr){5-6}\cmidrule(lr){7-8}
    \textbf{System} & total / active & WER $\downarrow$ & Sim $\uparrow$ & WER $\downarrow$ & Sim $\uparrow$ & WER $\downarrow$ & Sim $\uparrow$ \\
    \midrule
    Ground truth & -- & 9.66 & 70.7 & 9.19 & 79.3 & 9.49 & 73.8 \\
    \midrule
    \multicolumn{8}{l}{\emph{With reference transcript}} \\
    Qwen3-TTS 1.7B~\citep{hu2026qwen3ttstechnicalreport} & 1.7B & 22.80 & 80.6 & 14.61 & 74.5 & 19.82 & 78.4 \\
    Fish S2 Pro~\citep{liao2026fishaudios2technical} & 4.4B & 11.03 & 81.2 & 5.98 & 70.3 & 9.20 & 77.2 \\
    VoxCPM2~\citep{voxcpm2_2026} & 2B & 12.47 & \underline{86.2} & 5.12 & \textbf{77.8} & 9.80 & \textbf{83.1} \\
    CosyVoice 3~\citep{du2025cosyvoice} & 0.5B & 54.85 & 76.7 & 42.91 & 73.0 & 50.51 & 75.3 \\
    Higgs Audio v3~\citep{bosonai2026higgsaudiov3} & 4B & 9.42 & 79.5 & 12.66 & 69.2 & 10.60 & 75.8 \\
    OmniVoice~\citep{zhu2026omnivoice} & 0.6B & \underline{7.52} & \textbf{87.6} & \underline{5.03} & \underline{75.0} & \underline{6.62} & \underline{83.1} \\
    Audio8-TTS~\citep{audio8tts2026} & 0.6B & 13.45 & 80.3 & 7.33 & 67.5 & 11.22 & 75.6 \\
    Fish Audio Drama 3~\citep{han2026drama3} & 35B / 3B & \textbf{5.62} & 82.1 & \textbf{3.20} & 72.5 & \textbf{4.74} & 78.6 \\
    \midrule
    \multicolumn{8}{l}{\emph{Without reference transcript}} \\
    ZONOS2 8B~\citep{clark2026zonos2} & 8B / 0.9B & \textbf{7.10} & 75.0 & \textbf{4.68} & 57.8 & \textbf{6.22} & 68.8 \\
    Qwen3-TTS 1.7B~\citep{hu2026qwen3ttstechnicalreport} & 1.7B & 24.46 & 69.6 & 11.71 & 67.1 & 19.83 & 68.7 \\
    Qwen3-TTS 0.6B~\citep{hu2026qwen3ttstechnicalreport} & 0.6B & 30.39 & 69.2 & 15.07 & 64.3 & 24.82 & 67.4 \\
    VoxCPM2~\citep{voxcpm2_2026} & 2B & 13.72 & 78.9 & 5.25 & \underline{73.9} & 10.64 & 77.1 \\
    MOSS-TTS v1.5~\citep{gong2026mosstts} & 8B & 10.95 & \underline{82.9} & 7.22 & 72.0 & 9.60 & \underline{78.9} \\
    VibeVoice 7B~\citep{microsoft2025vibevoice} & 7B & 11.50 & 82.2 & 13.51 & 65.2 & 12.24 & 76.1 \\
    \midrule
    \zyoursrow ZONOS3-Mini + PT, embedding (ours) & 886M / 100M & 7.79 & \textbf{83.4} & \underline{4.81} & \textbf{77.3} & 6.71 & \textbf{81.2} \\
    \zyoursrow ZONOS3-Mini + PT, raw audio (ours) & 886M / 100M & \underline{7.39} & 76.5 & 5.48 & 58.1 & \underline{6.70} & 69.8 \\
    \bottomrule
  \end{tabular}
  \caption{Comparison with external systems on all ZTTS2-Eval languages (35 clean and 20
    in-the-wild parts). Systems are grouped by
    whether they are conditioned on the transcript of the reference clip. Best in bold, second best
    underlined, within each group. Per-language results for every system are given in Appendix~\ref{sec:perlang}.}
  \label{tab:external}
\end{table}

\Cref{tab:external} compares ZONOS3-Mini with other \ac{TTS} systems. Listening samples for all systems are available online.\footnote{\url{https://zyphra.github.io/ZTTS2-Eval/}}
Among systems that do not use the reference transcript, ZONOS3-Mini has the highest similarity (81.2 with embeddings) and the second-lowest macro \ac{WER} after ZONOS2 (6.70 with raw audio, 6.71 with embeddings).
Across all systems, the lowest \ac{WER} is achieved by Fish Audio Drama 3 (4.74), a 35B model with 3B active parameters,
followed by ZONOS2 (6.22) and OmniVoice (6.62). 
The in-the-wild similarity of the embedding path (77.3) is second only to VoxCPM2.
In macro similarity,
only VoxCPM2 and OmniVoice score higher than ZONOS3-Mini. We evaluate every system on every language. Qwen3-TTS and CosyVoice 3 officially support a smaller set of languages, which significantly affects their scores here. 
The raw-audio path of ZONOS3-Mini has a higher similarity
than ZONOS2 (69.8 against 68.8) and a higher macro \ac{WER} (6.70 against 6.22) with roughly nine times fewer parameters.

\begin{table}[t]
  \centering
  \footnotesize
  \setlength{\tabcolsep}{4pt}
  \begin{tabular}{llrrrr}
    \toprule
    & & \multicolumn{2}{c}{\makecell{\textbf{Clean,}\\\textbf{4 languages}}} & \multicolumn{2}{c}{\textbf{In the wild}} \\
    \cmidrule(lr){3-4}\cmidrule(lr){5-6}
    \textbf{System} & \textbf{Voice} & WER $\downarrow$ & Sim $\uparrow$ & WER $\downarrow$ & Sim $\uparrow$ \\
    \midrule
    Cartesia Sonic 3.6 & clone & 2.49 & \textbf{85.0} & \underline{3.30} & \underline{76.4} \\
    Inworld TTS 2 & clone & \underline{2.42} & 71.5 & 3.89 & 61.4 \\
    ElevenLabs v4 & clone & \textbf{2.07} & 82.9 & \textbf{2.97} & -- \\
    Gemini 3.8 Flash & preset & 2.55 & -- & 4.67 & -- \\
    \midrule
    \zyoursrow ZONOS3-Mini + PT, embedding (ours) & clone & 2.55 & \underline{83.1} & 5.00 & \textbf{77.5} \\
    \zyoursrow ZONOS3-Mini + PT, raw audio (ours) & clone & 2.56 & 75.5 & 5.86 & 58.1 \\
    \bottomrule
  \end{tabular}
  \caption{Comparison with closed-source systems on ZTTS2-Eval Mini, a subset of ZTTS2-Eval that reduces evaluation cost, restricted to the items every system generated. Best in bold, second best underlined.}
  \label{tab:closed_common}
\end{table}
\Cref{tab:closed_common} compares ZONOS3-Mini with closed-source systems on a subset (German, English, Spanish and Chinese) of ZTTS2-Eval. ZONOS3-Mini performs competitively, achieving the second-highest similarity on the clean set and the highest similarity on the in-the-wild set, while remaining close to the closed-source systems in terms of \ac{WER}.

\section{Ablations}\label{sect:arch}

\subsection{Backbone}\label{sec:backbone}
We selected the architecture 
from downstream performance on decoded audio, using
the evaluation sets in \Cref{sect:eval}.

\Cref{tab:ladder150k} reports a cumulative ablation of the backbone components on the
ZONOS2-style 800M/80M-active backbone at 150k steps.
Latent experts alone increased in-the-wild \ac{WER}. A linear router in place of the
\ac{MLP} router with \ac{EDA}~\citep{washbourne2026zaya18btechnicalreport} reduced it below the baseline. Quantile balancing gave the
highest in-the-wild similarity. Attention residuals gave the lowest \ac{WER} on both
sets. ZONOS3-Mini uses all four changes. 

\begin{table}[t]
  \centering
  \small
  \setlength{\tabcolsep}{6pt}
  \begin{tabular}{lrrrrrr}
    \toprule
    & \multicolumn{3}{c}{\textbf{In the wild}} & \multicolumn{3}{c}{\textbf{Clean}} \\
    \cmidrule(lr){2-4}\cmidrule(lr){5-7}
    \textbf{Arm (cumulative)} & WER $\downarrow$ & Sim $\uparrow$ & UTMOS $\uparrow$ & WER $\downarrow$ & Sim $\uparrow$ & UTMOS $\uparrow$ \\
    \midrule
    Original (ZONOS2 block) & 64.6 & 45.1 & 2.10 & 39.1 & 59.2 & 2.86 \\
    Latent \ac{MoE} & 83.2 & 44.2 & 1.95 & 33.7 & \textbf{69.3} & 2.87 \\
    $+$ linear router, no \ac{EDA} & 43.5 & 54.5 & 2.19 & 26.9 & 65.8 & \textbf{2.99} \\
    $+$ quantile balancing & 39.0 & \textbf{56.3} & \textbf{2.20} & 33.7 & 60.0 & 2.93 \\
    $+$ XSA & 49.1 & 53.1 & 2.19 & 31.2 & 67.0 & 2.95 \\
    $+$ AttnRes-4 (no XSA) & \textbf{36.3} & 55.5 & 2.18 & \textbf{22.9} &
    68.0 & 2.95 \\
    \bottomrule
  \end{tabular}
  \caption{150k backbone ablation on the four-language ZTTS2-Eval sets. Best in bold.}
  \label{tab:ladder150k}
\end{table}

\ac{XSA}~\citep{zhai2026xsa} removes each token's own value direction from the attention output, which improves text language model performance. In our setting it matched the baseline training loss and increased in-the-wild \ac{WER} by 10 points (\Cref{tab:ladder150k}). We attribute this to the delay pattern, in which the current position contains information about previous timesteps which is needed for the prediction of the current timestep.

We selected the 4.5:1 \ac{GDN}/attention layout from quality and inference-cost comparisons. On an 800M latent-\ac{MoE} backbone at 50k steps with all other settings fixed, the 4.5:1 \ac{GDN} layout matched softmax attention in every layer to within 0.1 \ac{WER} and 0.1 similarity on the in-the-wild set, while \ac{GLA}~\citep{yang2024gla} and \ac{SWA} scored 0.5 to 0.9 \ac{WER} points worse. On one H100 with CUDA graphs at 16k context, the 4.5:1 layout is 1.22$\times$ faster than global attention at batch 1 and 3.2$\times$ faster at batch 256, with 4.6\,GiB of persistent state at batch 256 instead of 22\,GiB.

\subsection{Codec Ablations}
\begin{table}[t]
  \centering
  \footnotesize
  \setlength{\tabcolsep}{4.5pt}
  \begin{tabular}{llrrlrrr}
    \toprule
    \textbf{Configuration} & \textbf{Family} & \textbf{Frames/s} & \textbf{Codebooks} & \textbf{Codebook sizes} & \textbf{kbit/s} & \textbf{\acs{WER}} $\downarrow$ & \textbf{Sim} $\uparrow$ \\
    \midrule
    Fish S1 codec~\citep{liao2026fishaudios2technical} & S1--\acs{DAC} & 21.53 & 10 & $1{\times}4096+9{\times}1024$ & 2.20 & \textbf{4.79} & \textbf{81.87} \\
  \textbf{ZAC-1} & dual hybrid & 15.00 & 10 & $10{\times}4096$ & 1.80 & 4.80 & 79.94 \\
    Dual hybrid, 10\,s & dual hybrid & 15.00 & 10 & $10{\times}4096$ & 1.80 & 5.81 & 74.54 \\
    Dual hybrid R1024 & dual hybrid & 15.00 & 10 & $1{\times}4096+9{\times}1024$ & 1.53 & 6.53 & 67.83 \\
    Hybrid, 15\,Hz, 10 CB & hybrid & 15.00 & 10 & $10{\times}4096$ & 1.80 & 7.25 & 68.37 \\
    Hybrid, 12.5\,Hz, 10 CB & hybrid & 12.50 & 10 & $10{\times}4096$ & 1.50 & 7.22 & 65.11 \\
    S1 retrain & S1--\acs{DAC} arch. & 21.53 & 10 & $10{\times}4096$ & 2.58 & 11.07 & 78.64 \\
    S1 MoE retrain & S1--\acs{DAC} + \acs{MoE} & 21.53 & 10 & $10{\times}4096$ & 2.58 & 12.77 & 77.67 \\
    MOSS-style, 10 CB & MOSS-v2 & 15.00 & 10 & $10{\times}4096$ & 1.80 & 10.46 & 66.11 \\
    Hybrid, 15\,Hz, 16 CB & hybrid & 15.00 & 16 & $16{\times}4096$ & 2.88 & 12.47 & 71.67 \\
    MOSS-style, 16 CB & MOSS-v2 & 15.00 & 16 & $16{\times}4096$ & 2.88 & 14.10 & 65.10 \\
    Hybrid, 12.5\,Hz, 16 CB & hybrid & 12.50 & 16 & $16{\times}4096$ & 2.40 & 16.87 & 51.82 \\
    \bottomrule
  \end{tabular}
  \caption{Downstream codec ablation.}
  \label{tab:codec_ablation}
\end{table}

\Cref{tab:codec_ablation} lists the codec variants trained on the way to ZAC-1, each scored after 50k steps of downstream \ac{TTS} training. Our final ZAC-1 matches the \ac{WER} of the Fish S1 codec~\citep{liao2026fishaudios2technical} at a lower frame rate and bitrate, with lower speaker similarity.

\subsection{Text tokenizer}\label{sec:tok_ablation}
\begin{figure}[t]
  \centering
  \includegraphics[width=\textwidth]{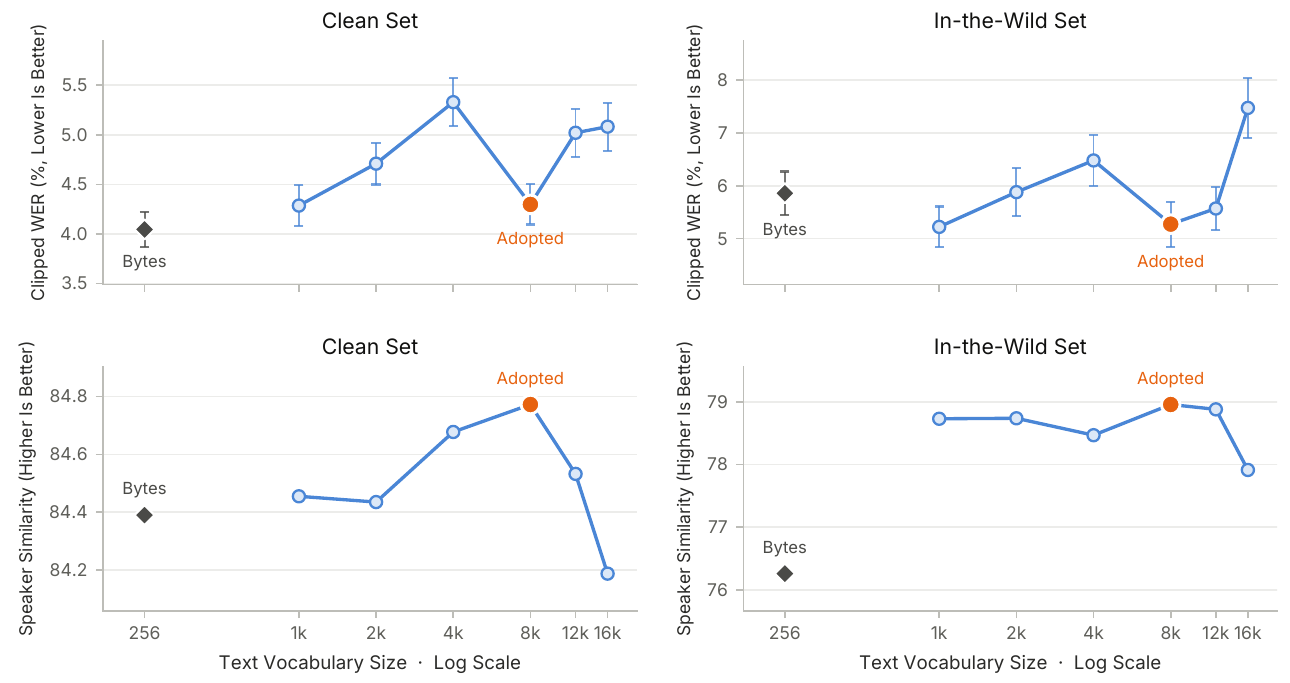}
  \caption{SuperBPE vocabulary sweep: matched 50k-step runs on the S1--DAC codec scored on the ZTTS2-Eval
    clean and in-the-wild sets (four-language macro over en, zh, es and ja; bars are 95\,\%
    confidence intervals of the clipped \ac{WER}).}
  \label{fig:tokenizer_sweep}
\end{figure}

We trained six 50k-step runs on the S1--DAC codec while sweeping SuperBPE vocabularies from 1,024 to 16,384 entries (\Cref{fig:tokenizer_sweep}). The 8,192-entry vocabulary compresses text to 3.49 bytes per token, beats the byte baseline on in-the-wild \ac{WER} and is best on in the wild and clean sets on speaker similarity; 16,384 is worst in the wild, driven by Japanese. We therefore kept 8,192.

\section{Inference}\label{sect:inference}
ZONOS3-Mini is served with SGLang-Omni~\citep{sglangomni2026}.
The backbone decode step runs in CUDA graphs and the codec decoder streams audio as
frames are generated. On one H100 with embedding prompts, the first audio arrives 85\,ms
after a single request and 80\,ms (p50) at 8 concurrent requests. Throughput reaches
187 seconds of audio per second at 8 concurrent requests and peaks at 457 at 64
(\Cref{fig:headline}).
Speaker \ac{CFG} runs a twin sequence without the reference prefix or speaker row and
guides the logits as $\ell_\mathrm{u}+S(\ell_\mathrm{c}-\ell_\mathrm{u})$ before
sampling, with a default scale of 1.5 for embedding prompts.

We experiment with low-bit weights on the 375k checkpoint, as shown in \Cref{tab:quant}. Each quantized
variant starts from BF16. The \ac{QAT} variants use a short distillation run with
\ac{KLD} against the BF16 model. Four-bit experts with 8-bit mixers and \ac{QAT}
are the smallest tested format without a measurable loss of intelligibility. The same format with \ac{PTQ} instead of \ac{QAT}, learned-clip
2-bit experts~\citep[\zypin{\S3.2.2}{2502.02631v2}{6}]{liu2025paretoq} and ternary experts trained with
BitDistill~\citep[\zypin{\S3}{2510.13998v1}{3}]{wu2025bitdistill} all reduce intelligibility. FP8 W8A8
weights~\citep{micikevicius2022fp8} increase throughput by at most 3\,\%.

\begin{table}[t]
  \centering
  \scriptsize
  \setlength{\tabcolsep}{3pt}
  \zycirange[2.2cm]{-0.5}{3.5}%
  \begin{tabular}{lrrrrrccc}
    \toprule
    \textbf{Arm} & \textbf{en} & \textbf{zh} & \textbf{ja} & \textbf{ko} & \textbf{all} & \textbf{$\Delta$ vs BF16 [95\,\% CI]} & \zyciaxis{0,1,2,3} & \textbf{Sim prompt / other} \\
    \midrule
    BF16 & 3.16 & 4.57 & 4.81 & \textbf{5.86} & \textbf{4.14} & -- & & 57.99 / 52.16 \\
    Q8\_0 mixers, \ac{PTQ} & 3.20 & 4.64 & 5.24 & 5.97 & 4.27 & $+0.14$ [$-0.23$, $+0.49$] & \zyci{+0.14}{-0.23}{+0.49} & 57.91 / 52.03 \\
    Q8\_0 mixers, \ac{QAT} & 3.24 & 4.93 & \textbf{4.74} & 6.28 & 4.29 & $+0.15$ [$-0.23$, $+0.51$] & \zyci{+0.15}{-0.23}{+0.51} & 57.97 / 52.01 \\
    Q4\_0 exp.\ + Q8\_0 mix., \ac{PTQ} & 3.71 & 4.89 & 4.95 & 7.45 & 4.68 & $+0.54$ [$+0.14$, $+0.93$] & \zyci{+0.54}{+0.14}{+0.93} & 57.47 / 51.70 \\
    Q4\_0 exp.\ + Q8\_0 mix., \ac{QAT} & \textbf{3.07} & \textbf{3.94} & 5.29 & 7.02 & 4.21 & $+0.07$ [$-0.34$, $+0.48$] & \zyci{+0.07}{-0.34}{+0.48} & 57.75 / 51.88 \\
    2-bit experts (learned clip), \ac{QAT} & 3.61 & 5.24 & 5.38 & 8.85 & 4.97 & $+0.83$ [$+0.43$, $+1.24$] & \zyci{+0.83}{+0.43}{+1.24} & 56.24 / 50.71 \\
    Ternary exp.\ + Q8\_0 mix., BitDistill & 4.42 & 6.50 & 8.76 & 14.78 & 7.05 & $+2.91$ [$+2.38$, $+3.43$] & \zyci{+2.91}{+2.38}{+3.43} & 54.73 / 49.23 \\
    FP8 W8A8, \ac{PTQ} & 3.72 & 5.40 & 6.10 & 8.02 & 5.11 & $+0.97$ [$+0.58$, $+1.36$] & \zyci{+0.97}{+0.58}{+1.36} & 57.05 / 51.28 \\
    FP8 W8A8, \ac{QAT} & 3.19 & 4.87 & 5.73 & 6.94 & 4.54 & $+0.40$ [$0.00$, $+0.81$] & \zyci{+0.40}{-0.00}{+0.81} & 57.54 / 51.62 \\
    \bottomrule
  \end{tabular}
  \caption{Quantization of the 375k pre-train on the four-language in-the-wild set at \ac{CFG} 1:
    clipped \ac{WER}, paired bootstrap difference to BF16, and similarity to the prompt and
    to another recording of the speaker. \ac{QAT} arms are distilled from the BF16 model. The \textbf{all} column scores every language of the set together rather than averaging the four columns shown. Lowest \ac{WER} per column in bold.}
  \label{tab:quant}
\end{table}

A native C++/ggml runtime~\citep{gerganov2023ggml} is released which runs the entire inference pipeline, including both \ac{iSTFT} heads and the speaker encoder. All tested formats meet internal quality thresholds for \ac{WER}, similarity, MSR-UTMOS and
failure rate. These range from Q8\_0 (1.03\,GB) to the 4-bit \ac{QAT} experts above
(0.66\,GB). On an H100 the quantized profiles decode about 20\,s of audio
per second for one stream, and on CPU 129 to 160 frames per second.  We also release a WebGPU~\citep{w3cwebgpu2026} runtime for in-browser inference. On a Pixel 9 and an M3 MacBook Air, the runtime generates speech in real time, with time to first audio below 800 ms and 450 ms, respectively.

\section{Adapting ZONOS3-Mini}\label{sect:adapt}
The adaptations below retain the backbone architecture. Voice design starts from the
final post-trained checkpoint, and the other adaptations start from the 375k pre-train. Each uses a separate evaluation set, specified below.

Voice design applies control directions to the speaker embedding to modify perceived age, gender presentation and vocal texture. Multiple directions can be combined to compose a target voice. Emotion and delivery control uses training-free activation steering~\citep{wang2026cocoemo}.
We compute mean differences between
Expresso~\citep{nguyen2023expresso} read and conversational styles at the output of the
layer-9 \ac{GDN} mixer. These are combined into six axes: valence, arousal,
dominance, loudness, tempo and clarity. For utterance-level steering,  the chosen vector is scaled and added to every generated frame at that layer. For word-level steering, alignment attention heads track the target word during streaming generation, and the steering intervention is restricted to its aligned span. For each voice, we calibrate both ends
of each axis with a threshold on the ParaSpeechCLAP similarity gain~\citep{diwan2026paraspeechclap}
and an Expresso emotion classifier. On a held-out set of generated audio, the three
default voices passed these checks for most of the 12 poles. The positive delivery
poles were disabled most often.

Accent control uses a separate \ac{LoRA} adapter~\citep{hu2021lora} for each of
14 accents. Each adapter trains on native-language speech from the fine-tuning subset,
presented as English~\citep[\zypin{\S3}{2603.07534v1}{3}]{lertpetchpun2026accent}. Its weight $\alpha$ sets the strength
of the adaptation at inference. Evaluation uses 200 English in-the-wild prompts per accent, with
Vox-Profile~\citep{feng2025voxprofile} for target-accent top-1 rate and Whisper for
\ac{WER}. Rank-64
\ac{LoRA} (8.9M parameters) increases the mean target-accent rate from 0.126 to 0.393 at
$\alpha{=}1$, and \ac{WER} from 3.71 to 14.26. At $\alpha{=}0.7$, every adapter has a
\ac{WER} near 6\,\%.

For speech editing, we train the 375k checkpoint for 10k steps on the pre-training mix
with VoiceCraft-style masked-span infilling~\citep[\zypin{\S3.1}{2403.16973v3}{3}]{peng2024voicecraft}. We remove one to three random spans identified by MMS forced alignments~\citep{pratap2023mms} from the target.
We append their codes after the context as an infill block with a dedicated span-end ID. On
Common Voice English (500 texts, 460 edit-eligible) scored with MMS greedy \ac{CTC},
edited utterances reach 13.60 \ac{WER}, compared with 10.17 for a resynthesis of the
original codes, and plain \ac{TTS} \ac{WER} increases by 0.8 points.

Multi-turn conversation uses CSM-style windows~\citep{sesame2025csm}, where the audio
and text of prior turns provide context for the target turn. The 375k pre-train
receives 25k further training steps. Evaluation uses 252
held-out windows from 45 English episodes and capped Whisper \ac{WER}.
The serving variant reaches 10.7\,\%, compared with 16.8\,\% for real recordings
under the same protocol.

\section{Discussion}\label{sect:discuss}

\subsection{Loss and quality}
Training loss alone did not reliably distinguish the better speech models.
Across 75 matched 400M-total/40M-active runs, the correlation between training \ac{NLL} and decoded quality was 0.52. The \ac{XSA} experiment gives a concrete example: loss stayed
unchanged while \ac{WER} rose by ten points (\Cref{sect:arch}). We therefore used
loss to monitor training progress and selected models from decoded audio scored
with the eight-part ZTTS2-Eval protocol.

\subsection{\texorpdfstring{\acs{MoE}}{MoE} for audio}
ZONOS2 found expert balancing less stable on audio than on text and used dense boundary layers and a manually tuned bias to stabilize it. ZONOS2 also found expert balancing to be well correlated with model stability, and in ZONOS3-Mini we observe the same. After adopting a linear router and \ac{QB}, we saw a significant drop in \ac{WER}. The 375k-step pre-train required no auxiliary loss, balancing adjustments, or dense boundary layers beyond the first.

\subsection{Reference clips versus embeddings}
When deciding between embedding cloning and reference-clip cloning, we weighed trade-offs in inference efficiency, clone quality, and clone controllability. MECT-B2 embeddings provide a controllable latent space for manipulating speaker characteristics such as age, gender and voice pitch, while providing a compressed representation of the clone audio. Frozen speaker embeddings also let us briefly train with the target audio as the reference, which improves similarity scores in evaluation. We chose to add encoded audio references to experiment with a higher-bandwidth channel to the reference and a solution that scales with training FLOPs and model size. Based on early evidence in ZONOS3-Mini we expect strong results when using audio references in larger models and foresee potential problems with frozen speaker embeddings on target audio.

\section{Conclusion}
For its inference speed and size, ZONOS3-Mini is a state-of-the-art \ac{TTS} system, and it can run anywhere. It combines a 15\,Hz ZAC-1 codec with a hybrid \ac{GDN}/attention
backbone, latent experts, attention residuals and SuperBPE text. Pre-training uses
same-recording raw-audio references; fine-tuning adds speaker embeddings and control
rows; post-training increases reliability and consistency, especially on difficult zero-shot prompt audio. On all ZTTS2-Eval
languages, ZONOS3-Mini performs favorably versus larger and slower state-of-the-art systems. Experiments also support 4-bit expert weights, emotion and accent
controls, speech editing and conversational generation. All components of the system are released open source, including inference and training code.\footnote{\url{https://github.com/Zyphra/ZONOS3-Mini}}

\begin{acknowledgements}
We thank Rishi Iyer for discussions on \ac{MoE} training dynamics and Nathan Kolbas for assistance in setting up and maintaining the ZONOS3-Mini model endpoint on Zyphra Cloud.
\end{acknowledgements}

\FloatBarrier
\bibliographystyle{zyphra/zyphra}
\bibliography{main}

\zyappendix

\section{Training hyper-parameters}
\begin{table}[H]
  \centering
  \scriptsize
  \begin{tabularx}{\linewidth}{l>{\raggedright\arraybackslash}X>{\raggedright\arraybackslash}X>{\raggedright\arraybackslash}X}
    \toprule
    & \textbf{Pre-train 375k} & \textbf{Fine-tune 60k} & \textbf{Rate stage 20k} \\
    \midrule
    Nodes $\times$ GPUs & 14 $\times$ 8 H100 & 14 $\times$ 8 H100 & 14 $\times$ 8 H100 \\
    Positions / GPU / step & 106,496 & 106,496 & 106,496 \\
    Positions / step & 11.93M & 11.93M & 11.93M \\
    Steps & 375,000 & 60,000 & 20,000 \\
    Muon / AdamW / bias LR & 0.02 / 0.002 / 0.002 & 0.0035 / 0.00035 / 0.00035 & 0.0035 / 0.00035 / 0.00035 \\
    Warm-up / schedule & 250 / cosine to $10^{-6}$ & 100 / cosine to 0 & 100 / cosine to 0 \\
    Weight decay / clip & 0.1 / 1.0 & 0.1 / 1.0 & 0.1 / 1.0 \\
    Precision / parallelism & BF16, FSDP full shard & same & same \\
    Data & pre-training corpus evenly weighted by hours & reference branch 45\,\%,
    embedding branch 45\,\%, no conditioning 10\,\% with CER agreement of 2\,\% across 3 transcripts & same \\
    Reference dropout & 0.1 & 0.1 & 0.1 \\
    Control-row dropout & -- & 0.2 each & 0.2 each \\
    Speaking-rate buckets & -- & 10 & 30 \\
    Speaker encoder & -- & MECT-B2 (trained in-house), 192-d & same \\
    Init & random & 375k checkpoint; fresh speaker projection and control tables & 60k checkpoint; rate table expanded to 30 buckets \\
    \bottomrule
  \end{tabularx}
  \caption{Training hyper-parameters of the released checkpoints.}
  \label{tab:hparams}
\end{table}

\section{Per-language results}\label{sec:perlang}
\Cref{tab:perlang_clean_a,tab:perlang_clean_b,tab:perlang_itw} give the per-language results behind
\Cref{tab:external} for every system: the 35 languages of the clean set and the 20 languages of the
in-the-wild set, with clipped \ac{WER} and speaker similarity for each.
\begin{table}[H]
  \centering
  \scriptsize
  \setlength{\tabcolsep}{2.4pt}
  \resizebox{\linewidth}{!}{%
  \begin{tabular}{llrrrrrrrrrrrrrrrrrr}
    \toprule
    \textbf{Model} & \textbf{Metric} & \textbf{en} & \textbf{zh} & \textbf{de} & \textbf{es} & \textbf{fr} & \textbf{it} & \textbf{ja} & \textbf{ko} & \textbf{ru} & \textbf{pt} & \textbf{ar} & \textbf{hi} & \textbf{tr} & \textbf{pl} & \textbf{id} & \textbf{th} & \textbf{fil} & \textbf{nl} \\
    \midrule
    \multicolumn{20}{l}{\emph{Reference}} \\
    \multirow{2}{*}{Ground truth} & WER $\downarrow$ & 4.30 & 3.02 & 4.64 & 2.89 & 4.08 & 4.05 & 4.16 & 3.20 & 5.67 & 4.18 & 13.69 & 14.61 & 9.60 & 4.48 & 7.31 & 5.44 & 13.43 & 6.83 \\
     & Spk.\ sim.\ $\uparrow$ & -- & -- & -- & -- & -- & -- & -- & -- & -- & -- & -- & -- & -- & -- & -- & -- & -- & -- \\
    \midrule
    \multicolumn{20}{l}{\emph{With reference transcript}} \\
    \multirow{2}{*}{Qwen3-TTS 1.7B~\citep{hu2026qwen3ttstechnicalreport}} & WER $\downarrow$ & 2.37 & 4.64 & 3.17 & 2.26 & 3.07 & 2.36 & 3.71 & 4.32 & 5.21 & 3.18 & 31.66 & 29.32 & 37.49 & 51.17 & 8.18 & 5.64 & 16.18 & 7.39 \\
     & Spk.\ sim.\ $\uparrow$ & 80.0 & 85.3 & 80.4 & 84.6 & 81.6 & 82.3 & 86.8 & 85.3 & 82.3 & 83.8 & 82.4 & 75.5 & 80.1 & 76.5 & 84.1 & 86.9 & 87.6 & 80.9 \\
    \multirow{2}{*}{Fish S2 Pro~\citep{liao2026fishaudios2technical}} & WER $\downarrow$ & 2.14 & 4.02 & 3.10 & 1.99 & 3.03 & 2.48 & 3.16 & 3.64 & 4.77 & 2.59 & 6.86 & 13.04 & 7.04 & 5.42 & 9.41 & 9.73 & 16.24 & 5.46 \\
     & Spk.\ sim.\ $\uparrow$ & 77.2 & 83.3 & 79.2 & 82.5 & 77.5 & 79.3 & 85.1 & 84.1 & 80.5 & 81.9 & 84.0 & 82.3 & 82.1 & 80.8 & 81.7 & 85.2 & 85.4 & 78.6 \\
    \multirow{2}{*}{VoxCPM2~\citep{voxcpm2_2026}} & WER $\downarrow$ & 3.93 & 4.63 & 4.06 & 3.21 & 4.33 & 3.52 & 4.44 & 4.20 & 5.43 & 3.38 & 8.00 & 10.27 & \underline{4.67} & 4.03 & 5.76 & \underline{4.40} & 9.25 & 5.28 \\
     & Spk.\ sim.\ $\uparrow$ & 84.0 & 88.2 & 86.0 & \underline{87.0} & \underline{84.7} & \underline{86.3} & 88.8 & 88.2 & 86.7 & \underline{86.9} & \textbf{87.7} & \underline{86.7} & \underline{86.4} & \underline{86.7} & \underline{88.6} & \underline{88.7} & \underline{89.4} & \underline{84.9} \\
    \multirow{2}{*}{CosyVoice 3~\citep{du2025cosyvoice}} & WER $\downarrow$ & 3.14 & 3.35 & 8.75 & 4.15 & 6.62 & 6.66 & 10.74 & 6.26 & 6.59 & 35.09 & -- & -- & 85.80 & 88.76 & 46.56 & -- & 44.50 & 53.84 \\
     & Spk.\ sim.\ $\uparrow$ & \textbf{85.2} & \textbf{90.5} & \underline{86.5} & \textbf{88.6} & \textbf{86.5} & 85.9 & \underline{89.1} & \underline{88.7} & \textbf{87.4} & 78.5 & -- & -- & 78.5 & 80.7 & 85.1 & -- & 87.2 & 78.2 \\
    \multirow{2}{*}{Higgs Audio v3~\citep{bosonai2026higgsaudiov3}} & WER $\downarrow$ & 3.56 & 6.29 & 5.20 & 3.18 & 4.74 & 4.27 & 5.08 & 5.69 & 6.48 & 4.75 & 11.65 & 13.78 & 8.30 & 5.81 & 5.91 & 6.65 & 10.78 & 8.05 \\
     & Spk.\ sim.\ $\uparrow$ & 72.4 & 81.0 & 77.1 & 79.6 & 75.0 & 77.1 & 81.9 & 81.8 & 79.5 & 80.8 & 79.7 & 79.3 & 79.4 & 78.3 & 82.1 & 83.7 & 84.2 & 75.4 \\
    \multirow{2}{*}{OmniVoice~\citep{zhu2026omnivoice}} & WER $\downarrow$ & 2.60 & \underline{3.34} & 3.40 & 2.13 & \underline{2.81} & 2.57 & \underline{3.10} & 3.44 & \underline{4.48} & 4.08 & 9.19 & 12.36 & 7.10 & \underline{3.49} & 8.32 & 5.90 & 10.96 & \underline{4.08} \\
     & Spk.\ sim.\ $\uparrow$ & \underline{84.6} & \underline{88.9} & \textbf{87.3} & 85.8 & 84.1 & \textbf{87.3} & \textbf{89.5} & \textbf{89.4} & \underline{87.2} & \textbf{88.2} & \underline{87.4} & \textbf{88.8} & \textbf{87.7} & \textbf{87.8} & \textbf{89.1} & \textbf{90.0} & \textbf{90.3} & \textbf{86.2} \\
    \multirow{2}{*}{Audio8-TTS~\citep{audio8tts2026}} & WER $\downarrow$ & 2.10 & 4.32 & 3.45 & 3.00 & 3.45 & 3.36 & 4.86 & 3.96 & 7.07 & 4.12 & 11.65 & 12.69 & 7.47 & 5.52 & 8.59 & 9.83 & 23.20 & 5.84 \\
     & Spk.\ sim.\ $\uparrow$ & 77.3 & 83.3 & 79.2 & 80.7 & 77.9 & 78.8 & 84.2 & 83.8 & 80.2 & 80.9 & 81.9 & 82.4 & 82.2 & 81.3 & 82.0 & 84.4 & 83.8 & 78.1 \\
    \multirow{2}{*}{Fish Audio Drama 3~\citep{han2026drama3}} & WER $\downarrow$ & \underline{1.92} & 3.52 & \textbf{2.54} & \textbf{1.75} & \textbf{2.56} & \textbf{1.50} & \textbf{2.61} & \underline{3.23} & \textbf{4.07} & \textbf{1.95} & \textbf{4.27} & \textbf{9.21} & \textbf{4.10} & \textbf{2.20} & 6.90 & \textbf{3.27} & \textbf{7.35} & \textbf{3.25} \\
     & Spk.\ sim.\ $\uparrow$ & 74.8 & 84.1 & 79.6 & 83.1 & 78.6 & 80.8 & 85.9 & 84.9 & 80.8 & 82.0 & 84.3 & 83.7 & 82.7 & 82.3 & 83.3 & 83.3 & 86.7 & 79.2 \\
    \midrule
    \multicolumn{20}{l}{\emph{Without reference transcript}} \\
    \multirow{2}{*}{ZONOS2 8B~\citep{clark2026zonos2}} & WER $\downarrow$ & 2.45 & 11.82 & 3.75 & 2.09 & 3.37 & \underline{1.94} & 7.20 & 5.90 & 4.80 & 2.87 & \underline{5.68} & \underline{9.97} & 5.32 & 3.77 & \textbf{3.89} & 4.93 & \underline{7.72} & 8.71 \\
     & Spk.\ sim.\ $\uparrow$ & 67.0 & 69.8 & 72.8 & 74.7 & 70.6 & 73.6 & 74.9 & 78.3 & 75.6 & 73.4 & 74.3 & 75.0 & 78.0 & 76.3 & 78.2 & 78.5 & 78.6 & 73.2 \\
    \multirow{2}{*}{Qwen3-TTS 1.7B~\citep{hu2026qwen3ttstechnicalreport}} & WER $\downarrow$ & \textbf{1.89} & \textbf{3.07} & \underline{2.66} & \underline{1.77} & 2.84 & 2.08 & 3.56 & \textbf{3.22} & 4.75 & 3.31 & 33.76 & 22.21 & 43.24 & 53.63 & 6.93 & 7.62 & 19.55 & 12.79 \\
     & Spk.\ sim.\ $\uparrow$ & 68.5 & 79.8 & 71.5 & 68.2 & 71.2 & 75.2 & 80.6 & 79.1 & 73.9 & 68.7 & 70.6 & 74.5 & 66.0 & 67.1 & 68.9 & 79.4 & 71.1 & 61.6 \\
    \multirow{2}{*}{Qwen3-TTS 0.6B~\citep{hu2026qwen3ttstechnicalreport}} & WER $\downarrow$ & 2.20 & 3.48 & 3.18 & 2.17 & 3.07 & 2.71 & 5.46 & 3.58 & 5.92 & 4.23 & 21.34 & 22.54 & 60.13 & 67.00 & 19.13 & 6.54 & 31.90 & 17.09 \\
     & Spk.\ sim.\ $\uparrow$ & 70.1 & 81.2 & 73.8 & 71.8 & 73.1 & 76.0 & 81.8 & 80.2 & 76.8 & 68.5 & 77.2 & 77.1 & 61.9 & 63.7 & 62.8 & 81.4 & 65.5 & 63.4 \\
    \multirow{2}{*}{VoxCPM2~\citep{voxcpm2_2026}} & WER $\downarrow$ & 3.59 & 5.06 & 4.50 & 3.70 & 5.10 & 3.94 & 5.06 & 4.00 & 5.84 & 4.44 & 9.28 & 10.17 & 5.89 & 4.86 & 5.82 & 5.35 & 8.92 & 6.53 \\
     & Spk.\ sim.\ $\uparrow$ & 66.1 & 77.0 & 80.6 & 74.8 & 76.4 & 80.6 & 82.9 & 83.2 & 80.5 & 80.1 & 80.5 & 81.8 & 81.5 & 80.6 & 84.0 & 83.8 & 85.2 & 78.5 \\
    \multirow{2}{*}{MOSS-TTS v1.5~\citep{gong2026mosstts}} & WER $\downarrow$ & 2.83 & 3.89 & 4.17 & 2.52 & 4.21 & 3.14 & 6.74 & 5.39 & 6.60 & 4.59 & 14.81 & 14.34 & 9.87 & 6.38 & 9.66 & 6.29 & 13.59 & 6.67 \\
     & Spk.\ sim.\ $\uparrow$ & 80.7 & 85.9 & 80.8 & 83.4 & 80.5 & 81.2 & 85.0 & 84.6 & 81.9 & 82.0 & 83.2 & 82.7 & 82.5 & 82.3 & 83.9 & 85.2 & 86.8 & 80.7 \\
    \multirow{2}{*}{VibeVoice 7B~\citep{microsoft2025vibevoice}} & WER $\downarrow$ & 2.81 & 3.68 & 5.81 & 3.02 & 6.73 & 4.45 & 10.59 & 5.58 & 6.84 & 4.96 & 15.78 & 13.26 & 10.16 & 8.64 & 7.26 & 7.52 & 12.45 & 9.63 \\
     & Spk.\ sim.\ $\uparrow$ & 79.9 & 84.5 & 78.8 & 82.3 & 77.4 & 80.2 & 84.8 & 85.4 & 82.7 & 85.0 & 80.5 & 83.8 & 83.3 & 80.6 & 83.8 & 85.5 & 87.4 & 78.7 \\
    \midrule
    \multicolumn{20}{l}{\emph{Ours}} \\
    \multirow{2}{*}{ZONOS3-Mini + PT, embedding (ours)} & WER $\downarrow$ & 2.34 & 4.13 & 3.56 & 2.13 & 3.49 & 2.43 & 3.71 & 4.99 & 5.61 & 2.59 & 12.76 & 12.48 & 6.51 & 3.74 & \underline{4.62} & 7.66 & 11.11 & 5.05 \\
     & Spk.\ sim.\ $\uparrow$ & 80.9 & 85.3 & 81.3 & 83.8 & 81.2 & 81.2 & 86.5 & 86.3 & 82.7 & 83.4 & 83.9 & 83.9 & 84.5 & 82.5 & 85.0 & 85.5 & 86.5 & 82.2 \\
    \multirow{2}{*}{ZONOS3-Mini + PT, raw audio (ours)} & WER $\downarrow$ & 2.16 & 4.10 & 3.80 & 1.99 & 4.01 & 2.10 & 3.99 & 5.28 & 5.39 & \underline{2.43} & 9.99 & 12.12 & 7.30 & 3.90 & 4.67 & 7.13 & 9.63 & 4.88 \\
     & Spk.\ sim.\ $\uparrow$ & 70.7 & 77.4 & 74.3 & 77.8 & 72.1 & 74.2 & 79.9 & 79.2 & 75.5 & 78.4 & 76.2 & 77.3 & 77.4 & 75.5 & 79.4 & 79.7 & 82.2 & 73.2 \\
    \bottomrule
  \end{tabular}}
  \caption{ZTTS2-Eval clean set (FLEURS-R) zero-shot results per language, part one of two: clipped \ac{WER} (\%, \ac{CER} for zh and ja) and ReDimNet speaker similarity (\texttimes 100). Systems are grouped by whether the harness gives them the reference transcript, as in \Cref{tab:external}. Best cloning system per language and metric in bold, second best underlined; the ground truth is shown for context and excluded from the ranking. A dash marks a language the system does not support.}
  \label{tab:perlang_clean_a}
\end{table}
\begin{table}[H]
  \centering
  \scriptsize
  \setlength{\tabcolsep}{2.4pt}
  \resizebox{\linewidth}{!}{%
  \begin{tabular}{llrrrrrrrrrrrrrrrrr}
    \toprule
    \textbf{Model} & \textbf{Metric} & \textbf{vi} & \textbf{cs} & \textbf{sv} & \textbf{da} & \textbf{fi} & \textbf{el} & \textbf{hu} & \textbf{ro} & \textbf{uk} & \textbf{nb} & \textbf{he} & \textbf{bg} & \textbf{hr} & \textbf{sk} & \textbf{ca} & \textbf{ms} & \textbf{fa} \\
    \midrule
    \multicolumn{19}{l}{\emph{Reference}} \\
    \multirow{2}{*}{Ground truth} & WER $\downarrow$ & 4.35 & 12.31 & 7.99 & 11.71 & 12.87 & 16.69 & 18.73 & 6.85 & 6.31 & 11.70 & 28.97 & 14.70 & 13.35 & 12.20 & 7.26 & 8.15 & 28.50 \\
     & Spk.\ sim.\ $\uparrow$ & -- & -- & -- & -- & -- & -- & -- & -- & -- & -- & -- & -- & -- & -- & -- & -- & -- \\
    \midrule
    \multicolumn{19}{l}{\emph{With reference transcript}} \\
    \multirow{2}{*}{Qwen3-TTS 1.7B~\citep{hu2026qwen3ttstechnicalreport}} & WER $\downarrow$ & 26.48 & 40.73 & 16.15 & 29.93 & 37.75 & 76.66 & 49.35 & 24.83 & 13.83 & 20.80 & 91.64 & 29.28 & 24.10 & 34.40 & 8.91 & 8.05 & 43.65 \\
     & Spk.\ sim.\ $\uparrow$ & 83.5 & 74.9 & 79.4 & 74.4 & 81.3 & 68.2 & 76.9 & 82.9 & 83.0 & 77.7 & 71.1 & 77.0 & 81.9 & 77.7 & 80.3 & 85.7 & 80.2 \\
    \multirow{2}{*}{Fish S2 Pro~\citep{liao2026fishaudios2technical}} & WER $\downarrow$ & 26.96 & 12.53 & 6.34 & 13.12 & 12.13 & 19.09 & 11.71 & 19.97 & 11.75 & 10.02 & 55.66 & 12.67 & 12.08 & 11.42 & 6.61 & 9.06 & 30.97 \\
     & Spk.\ sim.\ $\uparrow$ & 84.1 & 79.5 & 82.1 & 80.0 & 82.5 & 76.7 & 80.6 & 80.6 & 81.3 & 80.6 & 78.5 & 78.7 & 81.2 & 79.0 & 78.5 & 83.6 & 84.8 \\
    \multirow{2}{*}{VoxCPM2~\citep{voxcpm2_2026}} & WER $\downarrow$ & 5.09 & 32.04 & 6.34 & \underline{8.39} & 8.63 & \underline{8.90} & 64.19 & 29.92 & 12.98 & 8.26 & 14.55 & 32.26 & 24.18 & 30.90 & 14.64 & 6.82 & 35.75 \\
     & Spk.\ sim.\ $\uparrow$ & \underline{89.3} & 80.2 & \textbf{88.2} & \textbf{87.3} & \underline{88.0} & \underline{86.5} & 80.2 & \underline{86.2} & \textbf{86.7} & \textbf{87.0} & \underline{85.4} & 80.1 & 83.7 & 81.8 & 83.1 & \underline{89.1} & \underline{88.2} \\
    \multirow{2}{*}{CosyVoice 3~\citep{du2025cosyvoice}} & WER $\downarrow$ & -- & 83.31 & 55.14 & 64.59 & 66.03 & 99.84 & 81.70 & 61.41 & 91.11 & 54.44 & -- & 40.74 & 69.30 & 78.90 & 19.81 & 45.35 & -- \\
     & Spk.\ sim.\ $\uparrow$ & -- & 78.7 & 82.6 & 79.8 & 85.2 & 71.2 & 80.8 & 86.0 & 83.0 & 78.9 & -- & 81.0 & \underline{84.1} & \underline{82.2} & \underline{83.7} & 87.4 & -- \\
    \multirow{2}{*}{Higgs Audio v3~\citep{bosonai2026higgsaudiov3}} & WER $\downarrow$ & 5.58 & 12.12 & 7.86 & 12.54 & 13.24 & 14.07 & 14.27 & 7.36 & 7.26 & 11.95 & 21.46 & 12.12 & 12.68 & 13.21 & 7.22 & 7.93 & 28.57 \\
     & Spk.\ sim.\ $\uparrow$ & 83.7 & 79.4 & 80.8 & 79.9 & 80.7 & 80.0 & 78.9 & 81.5 & 79.9 & 78.9 & 75.7 & 76.6 & 79.5 & 77.5 & 75.8 & 81.7 & 82.9 \\
    \multirow{2}{*}{OmniVoice~\citep{zhu2026omnivoice}} & WER $\downarrow$ & \underline{3.43} & 10.75 & 10.42 & 8.50 & 7.95 & \textbf{8.11} & \underline{8.93} & \textbf{5.23} & \underline{5.57} & \underline{7.18} & 33.41 & \underline{7.88} & \underline{7.98} & 8.94 & \underline{3.90} & 9.88 & \underline{21.91} \\
     & Spk.\ sim.\ $\uparrow$ & \textbf{90.1} & \textbf{87.1} & \underline{87.0} & \underline{87.2} & \textbf{89.2} & \textbf{88.3} & \textbf{86.9} & \textbf{89.4} & \underline{84.8} & \underline{86.2} & \textbf{85.9} & \textbf{86.4} & \textbf{87.2} & \textbf{86.9} & \textbf{84.7} & \textbf{89.2} & \textbf{91.3} \\
    \multirow{2}{*}{Audio8-TTS~\citep{audio8tts2026}} & WER $\downarrow$ & 24.77 & 14.92 & 6.99 & 13.89 & 11.55 & 19.82 & 23.04 & 21.49 & 22.59 & 10.68 & 53.75 & 32.70 & 13.40 & 14.69 & 10.61 & 8.80 & 42.58 \\
     & Spk.\ sim.\ $\uparrow$ & 83.7 & 78.5 & 82.0 & 80.0 & 82.8 & 76.3 & 79.5 & 80.3 & 79.5 & 80.2 & 77.9 & 70.7 & 81.3 & 79.1 & 74.8 & 83.5 & 76.5 \\
    \multirow{2}{*}{Fish Audio Drama 3~\citep{han2026drama3}} & WER $\downarrow$ & \textbf{2.30} & \textbf{6.30} & \textbf{3.79} & \textbf{5.82} & \textbf{6.50} & 13.33 & \textbf{7.24} & 6.54 & \textbf{4.91} & \textbf{5.27} & 17.27 & \textbf{7.49} & \textbf{6.99} & \textbf{7.08} & \textbf{3.71} & 6.14 & \textbf{19.75} \\
     & Spk.\ sim.\ $\uparrow$ & 85.3 & 80.8 & 82.8 & 81.2 & 84.0 & 80.8 & 82.0 & 83.3 & 81.8 & 80.9 & 79.6 & 80.3 & 81.2 & 82.0 & 79.9 & 83.2 & 83.7 \\
    \midrule
    \multicolumn{19}{l}{\emph{Without reference transcript}} \\
    \multirow{2}{*}{ZONOS2 8B~\citep{clark2026zonos2}} & WER $\downarrow$ & 4.75 & \underline{6.85} & \underline{5.66} & 10.83 & \underline{6.63} & 11.10 & 11.87 & \underline{5.56} & 6.14 & 8.81 & \textbf{11.75} & 8.59 & 8.20 & \underline{8.56} & 7.20 & \textbf{4.87} & 24.96 \\
     & Spk.\ sim.\ $\uparrow$ & 77.6 & 75.0 & 76.8 & 75.8 & 78.1 & 76.2 & 74.8 & 78.3 & 75.3 & 75.4 & 69.7 & 74.1 & 72.7 & 74.0 & 72.1 & 76.4 & 79.2 \\
    \multirow{2}{*}{Qwen3-TTS 1.7B~\citep{hu2026qwen3ttstechnicalreport}} & WER $\downarrow$ & 68.55 & 45.03 & 19.07 & 27.16 & 50.62 & 60.71 & 51.70 & 27.63 & 14.29 & 22.89 & 88.29 & 28.31 & 25.55 & 37.07 & 10.83 & 7.67 & 41.97 \\
     & Spk.\ sim.\ $\uparrow$ & 68.4 & 63.5 & 64.8 & 63.0 & 62.5 & 65.5 & 65.8 & 69.8 & 75.8 & 57.6 & 70.6 & 66.0 & 71.6 & 69.7 & 65.5 & 71.9 & 66.5 \\
    \multirow{2}{*}{Qwen3-TTS 0.6B~\citep{hu2026qwen3ttstechnicalreport}} & WER $\downarrow$ & 63.44 & 59.94 & 28.68 & 32.71 & 66.15 & 77.97 & 66.49 & 36.91 & 22.98 & 32.69 & 87.34 & 31.46 & 47.36 & 56.42 & 9.86 & 18.93 & 42.60 \\
     & Spk.\ sim.\ $\uparrow$ & 70.6 & 60.8 & 63.1 & 63.1 & 60.2 & 67.2 & 62.4 & 65.5 & 75.7 & 57.4 & 70.4 & 64.9 & 66.1 & 63.6 & 69.7 & 65.5 & 68.7 \\
    \multirow{2}{*}{VoxCPM2~\citep{voxcpm2_2026}} & WER $\downarrow$ & 4.39 & 37.26 & 7.66 & 9.93 & 10.48 & 9.25 & 65.38 & 31.74 & 15.23 & 10.82 & 13.98 & 35.44 & 28.00 & 36.48 & 16.26 & 7.63 & 38.30 \\
     & Spk.\ sim.\ $\uparrow$ & 83.5 & 69.7 & 82.4 & 81.8 & 83.4 & 81.8 & 70.6 & 78.2 & 77.8 & 81.3 & 79.2 & 71.3 & 72.2 & 72.5 & 73.7 & 84.2 & 80.8 \\
    \multirow{2}{*}{MOSS-TTS v1.5~\citep{gong2026mosstts}} & WER $\downarrow$ & 10.91 & 13.47 & 8.97 & 12.61 & 15.28 & 12.51 & 16.65 & 8.31 & 12.33 & 12.34 & 31.99 & 17.54 & 20.59 & 16.41 & 10.20 & 8.98 & 28.54 \\
     & Spk.\ sim.\ $\uparrow$ & 83.4 & 82.0 & 84.3 & 82.8 & 84.4 & 82.5 & 81.0 & 84.0 & 82.5 & 82.8 & 80.1 & \underline{81.5} & 82.5 & 81.3 & 80.5 & 85.8 & 86.6 \\
    \multirow{2}{*}{VibeVoice 7B~\citep{microsoft2025vibevoice}} & WER $\downarrow$ & 16.76 & 14.84 & 11.48 & 21.09 & 20.27 & 11.83 & 19.29 & 11.02 & 10.63 & 17.84 & 15.44 & 13.94 & 16.22 & 16.42 & 9.94 & 8.99 & 27.50 \\
     & Spk.\ sim.\ $\uparrow$ & 84.6 & 80.0 & 83.0 & 81.0 & 83.4 & 81.0 & 81.3 & 83.7 & 82.2 & 80.7 & 79.0 & 80.5 & 81.4 & 80.9 & 79.4 & 84.9 & 86.8 \\
    \midrule
    \multicolumn{19}{l}{\emph{Ours}} \\
    \multirow{2}{*}{ZONOS3-Mini + PT, embedding (ours)} & WER $\downarrow$ & 4.54 & 9.23 & 7.54 & 10.05 & 10.31 & 10.98 & 11.13 & 7.25 & 6.96 & 9.66 & 15.09 & 9.73 & 10.31 & 11.17 & 5.69 & 6.63 & 27.50 \\
     & Spk.\ sim.\ $\uparrow$ & 86.8 & \underline{82.6} & 83.6 & 83.3 & 85.1 & 83.1 & \underline{82.5} & 84.3 & 82.7 & 81.7 & 80.6 & 81.0 & 81.6 & 81.3 & 81.6 & 86.1 & 86.2 \\
    \multirow{2}{*}{ZONOS3-Mini + PT, raw audio (ours)} & WER $\downarrow$ & 4.81 & 8.57 & 6.87 & 11.28 & 10.63 & 9.69 & 11.22 & 6.36 & 6.31 & 9.58 & \underline{13.25} & 9.04 & 9.40 & 10.00 & 4.68 & \underline{6.04} & 26.09 \\
     & Spk.\ sim.\ $\uparrow$ & 79.8 & 75.5 & 78.0 & 76.4 & 77.8 & 75.5 & 75.3 & 77.7 & 76.0 & 74.3 & 72.8 & 72.7 & 74.5 & 75.3 & 73.1 & 79.6 & 81.4 \\
    \bottomrule
  \end{tabular}}
  \caption{ZTTS2-Eval clean set (FLEURS-R) zero-shot results per language, part two of two: clipped \ac{WER} (\%, \ac{CER} for zh and ja) and ReDimNet speaker similarity (\texttimes 100). Systems are grouped by whether the harness gives them the reference transcript, as in \Cref{tab:external}. Best cloning system per language and metric in bold, second best underlined; the ground truth is shown for context and excluded from the ranking. A dash marks a language the system does not support.}
  \label{tab:perlang_clean_b}
\end{table}
\begin{table}[H]
  \centering
  \scriptsize
  \setlength{\tabcolsep}{2.4pt}
  \resizebox{\linewidth}{!}{%
  \begin{tabular}{llrrrrrrrrrrrrrrrrrrrr}
    \toprule
    \textbf{Model} & \textbf{Metric} & \textbf{en} & \textbf{zh} & \textbf{de} & \textbf{es} & \textbf{fr} & \textbf{it} & \textbf{ja} & \textbf{ko} & \textbf{ru} & \textbf{pt} & \textbf{ar} & \textbf{hi} & \textbf{id} & \textbf{tr} & \textbf{tl} & \textbf{pl} & \textbf{th} & \textbf{nl} & \textbf{uk} & \textbf{vi} \\
    \midrule
    \multicolumn{22}{l}{\emph{Reference}} \\
    \multirow{2}{*}{Ground truth} & WER $\downarrow$ & 3.82 & 8.78 & 6.25 & 5.08 & 5.57 & 6.25 & 6.46 & 9.02 & 9.53 & 5.58 & 11.62 & 13.34 & 15.18 & 17.24 & 12.54 & 9.90 & 8.74 & 6.51 & 12.20 & 10.28 \\
     & Spk.\ sim.\ $\uparrow$ & 76.3 & 80.6 & 77.9 & 78.9 & 77.7 & 79.1 & 80.4 & 79.0 & 79.2 & 80.1 & 79.8 & 78.0 & 78.8 & 80.2 & 78.3 & 80.2 & 81.5 & 76.1 & 80.8 & 82.2 \\
    \midrule
    \multicolumn{22}{l}{\emph{With reference transcript}} \\
    \multirow{2}{*}{Qwen3-TTS 1.7B~\citep{hu2026qwen3ttstechnicalreport}} & WER $\downarrow$ & 2.63 & 5.79 & 2.92 & 2.25 & 2.69 & 2.99 & 4.46 & 4.78 & 5.05 & 4.23 & 24.76 & 67.34 & 11.17 & 35.52 & 10.69 & 49.40 & 7.19 & 7.38 & 13.26 & 27.78 \\
     & Spk.\ sim.\ $\uparrow$ & 74.7 & \underline{79.4} & 76.2 & 76.4 & 75.5 & 76.5 & 78.7 & \textbf{78.8} & 78.7 & 76.3 & 75.3 & 45.5 & 76.3 & 73.5 & 73.5 & 70.4 & \underline{79.3} & 73.5 & 77.7 & 73.6 \\
    \multirow{2}{*}{Fish S2 Pro~\citep{liao2026fishaudios2technical}} & WER $\downarrow$ & 1.66 & \underline{2.36} & 2.24 & 1.76 & 2.39 & 2.45 & 1.85 & 2.58 & 3.53 & 2.42 & 7.14 & 10.28 & 11.04 & 7.19 & 10.73 & 3.74 & 5.70 & 4.74 & 8.92 & 26.95 \\
     & Spk.\ sim.\ $\uparrow$ & 65.3 & 73.8 & 68.9 & 68.8 & 67.6 & 69.8 & 72.4 & 72.6 & 72.0 & 70.7 & 71.9 & 69.1 & 67.5 & 72.9 & 63.0 & 72.9 & 75.1 & 67.2 & 71.6 & 72.7 \\
    \multirow{2}{*}{VoxCPM2~\citep{voxcpm2_2026}} & WER $\downarrow$ & 2.68 & 3.40 & 3.26 & 2.66 & 3.55 & 2.83 & 3.30 & 3.88 & 4.68 & 3.12 & 7.85 & 9.24 & 9.32 & 6.75 & 7.39 & 3.80 & 4.46 & 4.53 & 11.14 & 4.55 \\
     & Spk.\ sim.\ $\uparrow$ & 73.7 & 79.1 & \underline{77.0} & 76.8 & \underline{76.0} & \underline{78.0} & 78.0 & 77.8 & \underline{78.8} & \underline{77.8} & \textbf{79.0} & \textbf{76.6} & \textbf{77.2} & \textbf{79.3} & \textbf{75.1} & \textbf{80.3} & \textbf{80.0} & \underline{75.0} & \underline{79.6} & \textbf{80.9} \\
    \multirow{2}{*}{CosyVoice 3~\citep{du2025cosyvoice}} & WER $\downarrow$ & 2.68 & 3.64 & 4.12 & 2.81 & 4.41 & 3.93 & 7.49 & -- & 5.52 & 44.05 & -- & 99.10 & 51.16 & 85.72 & 22.43 & 88.59 & -- & 55.44 & 77.16 & 99.36 \\
     & Spk.\ sim.\ $\uparrow$ & \textbf{78.0} & \textbf{85.3} & \textbf{81.3} & \textbf{80.6} & \textbf{80.6} & \textbf{81.3} & \textbf{83.3} & -- & \textbf{82.8} & 70.1 & -- & 43.5 & 74.2 & 72.4 & 73.3 & 74.5 & -- & 69.9 & \textbf{79.7} & 67.9 \\
    \multirow{2}{*}{Higgs Audio v3~\citep{bosonai2026higgsaudiov3}} & WER $\downarrow$ & 6.57 & 8.68 & 9.18 & 8.73 & 8.69 & 7.85 & 11.79 & 12.87 & 9.96 & 8.47 & 22.36 & 20.57 & 16.74 & 18.66 & 16.04 & 11.29 & 14.68 & 11.30 & 13.21 & 15.50 \\
     & Spk.\ sim.\ $\uparrow$ & 64.8 & 70.9 & 68.0 & 68.8 & 66.6 & 69.4 & 69.8 & 68.9 & 70.5 & 71.0 & 69.0 & 66.1 & 69.0 & 71.2 & 66.1 & 72.6 & 72.2 & 66.8 & 70.9 & 72.1 \\
    \multirow{2}{*}{OmniVoice~\citep{zhu2026omnivoice}} & WER $\downarrow$ & 2.01 & 3.18 & 3.22 & 2.39 & 2.98 & 3.61 & 2.66 & 2.92 & 3.93 & 3.44 & 7.95 & 8.84 & 9.47 & 10.63 & 10.64 & 4.01 & 4.03 & 4.01 & 7.95 & \underline{2.72} \\
     & Spk.\ sim.\ $\uparrow$ & 69.9 & 76.7 & 74.8 & 72.4 & 71.3 & 75.2 & 75.3 & 77.1 & 76.2 & 74.4 & \underline{76.7} & \underline{74.9} & 73.5 & 77.2 & 70.0 & 79.0 & 78.3 & 73.1 & 75.6 & 78.8 \\
    \multirow{2}{*}{Audio8-TTS~\citep{audio8tts2026}} & WER $\downarrow$ & \underline{1.49} & 2.38 & 2.48 & 2.47 & 2.38 & 3.05 & 2.26 & 3.03 & 4.74 & 3.83 & 14.93 & 10.97 & 12.87 & 8.29 & 12.63 & 4.38 & 6.00 & 4.59 & 20.36 & 23.48 \\
     & Spk.\ sim.\ $\uparrow$ & 63.3 & 72.3 & 67.0 & 64.5 & 66.2 & 65.8 & 70.6 & 71.4 & 68.6 & 66.5 & 67.7 & 65.4 & 63.8 & 70.4 & 59.3 & 70.6 & 73.3 & 64.5 & 67.1 & 71.2 \\
    \multirow{2}{*}{Fish Audio Drama 3~\citep{han2026drama3}} & WER $\downarrow$ & \textbf{1.46} & 2.44 & \textbf{2.03} & 1.40 & \underline{1.85} & \textbf{1.55} & \textbf{1.59} & \textbf{2.28} & \underline{3.38} & \textbf{2.00} & \textbf{5.19} & \textbf{7.48} & \textbf{7.55} & \textbf{4.74} & \textbf{5.21} & \textbf{2.28} & \textbf{1.78} & \textbf{2.28} & \textbf{5.14} & \textbf{2.40} \\
     & Spk.\ sim.\ $\uparrow$ & 65.1 & 75.0 & 71.0 & 72.1 & 70.0 & 71.6 & 74.5 & 73.9 & 73.8 & 72.9 & 74.2 & 71.1 & 71.3 & 74.0 & 69.1 & 74.9 & 76.3 & 68.5 & 73.7 & 77.0 \\
    \midrule
    \multicolumn{22}{l}{\emph{Without reference transcript}} \\
    \multirow{2}{*}{ZONOS2 8B~\citep{clark2026zonos2}} & WER $\downarrow$ & 2.09 & 6.53 & 2.34 & \textbf{1.28} & 2.53 & 1.86 & 4.17 & 5.05 & 3.64 & 2.34 & \underline{7.12} & 8.87 & \underline{8.52} & \underline{6.12} & \underline{6.84} & 3.68 & 5.13 & 4.30 & 6.29 & 4.93 \\
     & Spk.\ sim.\ $\uparrow$ & 48.7 & 61.8 & 54.8 & 55.4 & 53.3 & 55.7 & 58.6 & 59.6 & 58.6 & 57.3 & 58.4 & 54.2 & 59.2 & 61.3 & 56.1 & 62.4 & 63.2 & 54.6 & 59.9 & 63.4 \\
    \multirow{2}{*}{Qwen3-TTS 1.7B~\citep{hu2026qwen3ttstechnicalreport}} & WER $\downarrow$ & 2.52 & \textbf{2.19} & \underline{2.05} & \underline{1.35} & \textbf{1.70} & \underline{1.66} & \underline{1.75} & \underline{2.39} & \textbf{3.15} & \underline{2.12} & 13.91 & 9.89 & 10.12 & 35.95 & 11.96 & 51.42 & 8.18 & 10.23 & 12.85 & 48.82 \\
     & Spk.\ sim.\ $\uparrow$ & 74.7 & 73.2 & 68.0 & 65.9 & 66.7 & 68.6 & 73.4 & 73.1 & 70.8 & 67.7 & 67.2 & 67.5 & 62.6 & 61.0 & 57.8 & 61.8 & 72.1 & 58.2 & 68.6 & 64.0 \\
    \multirow{2}{*}{Qwen3-TTS 0.6B~\citep{hu2026qwen3ttstechnicalreport}} & WER $\downarrow$ & 2.94 & 2.47 & 2.18 & 1.62 & 2.17 & 2.11 & 2.32 & 2.55 & 3.66 & 2.90 & 20.28 & 15.28 & 15.57 & 52.09 & 17.89 & 61.61 & 5.71 & 16.03 & 22.40 & 49.68 \\
     & Spk.\ sim.\ $\uparrow$ & 73.5 & 71.3 & 65.7 & 63.2 & 64.2 & 65.3 & 70.4 & 70.6 & 67.3 & 64.9 & 65.8 & 64.4 & 54.0 & 55.5 & 52.6 & 57.4 & 73.0 & 56.7 & 66.9 & 62.4 \\
    \multirow{2}{*}{VoxCPM2~\citep{voxcpm2_2026}} & WER $\downarrow$ & 2.17 & 2.94 & 3.73 & 2.41 & 3.39 & 2.65 & 4.65 & 3.74 & 4.66 & 3.98 & 9.89 & \underline{8.83} & 9.33 & 6.55 & 7.29 & 4.37 & 4.54 & 4.45 & 12.36 & 3.08 \\
     & Spk.\ sim.\ $\uparrow$ & 69.2 & 77.4 & 73.1 & 70.4 & 71.2 & 73.3 & 74.5 & 74.7 & 75.0 & 72.8 & 74.7 & 72.7 & 73.7 & 75.9 & 70.6 & 76.7 & 78.1 & 71.3 & 75.9 & 76.4 \\
    \multirow{2}{*}{MOSS-TTS v1.5~\citep{gong2026mosstts}} & WER $\downarrow$ & 3.06 & 3.50 & 4.08 & 3.00 & 4.88 & 4.23 & 4.86 & 4.69 & 6.22 & 4.82 & 19.15 & 11.59 & 14.30 & 10.18 & 10.43 & 6.18 & 3.60 & 5.26 & 13.91 & 6.54 \\
     & Spk.\ sim.\ $\uparrow$ & 69.7 & 76.2 & 71.4 & 71.2 & 70.3 & 72.5 & 72.3 & 72.7 & 73.4 & 71.6 & 71.3 & 69.3 & 70.2 & 73.5 & 68.1 & 74.7 & 74.9 & 68.6 & 73.7 & 74.1 \\
    \multirow{2}{*}{VibeVoice 7B~\citep{microsoft2025vibevoice}} & WER $\downarrow$ & 5.55 & 5.70 & 8.63 & 7.18 & 9.94 & 7.23 & 18.87 & 14.58 & 8.66 & 8.16 & 23.34 & 15.04 & 21.23 & 15.57 & 18.46 & 11.63 & 14.76 & 11.99 & 14.62 & 29.14 \\
     & Spk.\ sim.\ $\uparrow$ & 64.7 & 68.7 & 64.6 & 66.0 & 62.3 & 66.7 & 62.5 & 62.6 & 68.9 & 68.3 & 61.5 & 64.1 & 62.2 & 68.1 & 60.8 & 67.8 & 68.5 & 63.6 & 67.1 & 65.9 \\
    \midrule
    \multicolumn{22}{l}{\emph{Ours}} \\
    \multirow{2}{*}{ZONOS3-Mini + PT, embedding (ours)} & WER $\downarrow$ & 1.94 & 2.96 & 2.80 & 1.86 & 2.68 & 2.08 & 2.68 & 4.59 & 4.39 & 2.84 & 10.88 & 10.55 & 9.34 & 7.94 & 7.40 & \underline{3.40} & \underline{3.45} & \underline{3.86} & 7.03 & 3.53 \\
     & Spk.\ sim.\ $\uparrow$ & \underline{75.0} & 79.0 & 76.2 & \underline{77.0} & 75.7 & 77.6 & \underline{79.2} & \underline{78.7} & 77.8 & \textbf{78.0} & 76.3 & 73.9 & \underline{77.2} & \underline{78.1} & \underline{74.5} & \underline{79.1} & 79.2 & \textbf{75.4} & 78.1 & \underline{79.4} \\
    \multirow{2}{*}{ZONOS3-Mini + PT, raw audio (ours)} & WER $\downarrow$ & 1.99 & 4.29 & 2.76 & 2.40 & 2.85 & 2.55 & 4.13 & 10.76 & 4.65 & 2.78 & 10.23 & 9.75 & 9.29 & 7.71 & 7.84 & 4.28 & 6.11 & 3.88 & \underline{6.00} & 5.27 \\
     & Spk.\ sim.\ $\uparrow$ & 52.7 & 58.7 & 55.6 & 57.5 & 54.6 & 57.5 & 59.6 & 55.9 & 59.1 & 59.8 & 58.0 & 55.0 & 59.5 & 61.0 & 54.7 & 61.6 & 62.8 & 54.0 & 61.2 & 63.6 \\
    \bottomrule
  \end{tabular}}
  \caption{ZTTS2-Eval in-the-wild set (VoxBlink2) zero-shot results per language: clipped \ac{WER} (\%, \ac{CER} for zh and ja) and ReDimNet similarity to the prompt (\texttimes 100). Grouping and marking as in \Cref{tab:perlang_clean_a}.}
  \label{tab:perlang_itw}
\end{table}

\end{document}